\documentclass[12pt]{article}

\usepackage{mathrsfs}
\usepackage{fullpage}
\usepackage{amsfonts}
\usepackage{graphicx}
\usepackage{mathrsfs}
\usepackage{amsmath}
\usepackage{amssymb}
\usepackage{float}
\usepackage{subfig}
\usepackage{rotating,color}
\usepackage{xcolor,natbib,epstopdf}
\usepackage{appendix}
	\renewcommand\appendix{\par
	\setcounter{section}{0}
	\setcounter{subsection}{0}
	\setcounter{lemma}{0}
	\setcounter{equation}{0}
	\renewcommand{\thelemma}{\arabic{lemma}}
	\gdef\thesection{Appendix \Alph{section}}
	\gdef\thesubsection{\Alph{section}.\arabic{subsection}}}
\renewcommand\arraystretch{0.8}

\newcommand{\be}{\begin{equation}}
\newcommand{\ee}{\end{equation}}
\newcommand{\beaa}{\begin{eqnarray*}}
\newcommand{\eeaa}{\end{eqnarray*}}
\newcommand{\bea}{\begin{eqnarray}}
\newcommand{\eea}{\end{eqnarray}}

\newcommand{\eq}[1]{$(\ref{#1})$}

\newtheorem{theorem}{ \noindent T{\footnotesize HEOREM}}
\newtheorem{prop}{ \noindent P{\footnotesize ROPOSITION}}[section]
\newtheorem{lemma}{ \noindent L{\footnotesize EMMA}}[section]

\newtheorem{remark}{ \noindent R{\footnotesize EMARK}}[section]

\def\var{\mathrm {var}}

\def\diag{\mathrm {diag}}

\newcommand{\bm}{\boldsymbol}

\def\tr{\mathrm {tr}}
\def\U{{\bf U}}

\def\V{{\bm V}}
\def\M{{\bf M}}
\def\B{{\bf B}}

\def\R{{\bf R}}
\def\I{{\bf I}}
\def\h{{\bm h}}

\def\P{{\bf M}}
\def\Z{{\bm Z}}

\def\X{{\bf F}}

\def\Y{{\bm Y}}
\def\f{{\bm f}}

\def\L{{\bf \Lambda}}

\def\D{{\bf D}}

\def\tr{\mathrm {tr}}

\def\bms{{\bm\Sigma}}

\def\cp{\mathop{\rightarrow}\limits^{p}}
\def\cd{\mathop{\rightarrow}\limits^{d}}
\def\mR{\mathbb{R}}
\def\bma{\bm \alpha}
\def\bmv{\bm \varepsilon}

\def\boxit#1{\vbox{\hrule\hbox{\vrule\kern6pt  \vbox{\kern6pt#1\kern6pt}\kern6pt\vrule}\hrule}}
\def\bse{\begin{eqnarray*}}
\def\ese{\end{eqnarray*}}
\def\be{\begin{eqnarray}}
\def\ee{\end{eqnarray}}
\def\bsq{\begin{equation*}}
\def\esq{\end{equation*}}
\def\bq{\begin{equation}}
\def\eq{\end{equation}}

\def\var{\hbox{var}}

\def\mR{\mathbb{R}}
\def\n{\nonumber}

\def\diag{\mbox{diag}}
\def\tr{\mbox{tr}}

\def\trans{^\top}

\def\B{{\bf B}}

\def\D{{\bf D}}

\def\V{{\bf V}}

\def\F{{\bf F}}

\def\I{{\bf I}}
\def\M{{\bf M}}

\def\R{{\bf R}}

\def\U{{\bf U}}

\def\Y{{\bf Y}}

\def\Z{{\bf Z}}

\def\bSig{{\bf \Sigma}}

\def\diag{\hbox{diag}}

\def\diag{\hbox{diag}}

\def\squarebox#1{\hbox to #1{\hfill\vbox to #1{\vfill}}}

\def\0{{\bf 0}}
\def\1{{\bf 1}}

\def\mR{\mathcal R}

\def\var{\hbox{var}}

\def\wh{\widehat}
\def\wt{\widetilde}
\def\diag{\hbox{diag}}

\def\weps{\wt{\bm  \epsilon}}
\def\diag{\hbox{diag}}

\allowdisplaybreaks[4]

\title
{\bf Double Robust high dimensional alpha test for linear factor pricing model}
\author{Ping Zhao$^{1}$, Long Feng$^{1}$, Hongfei Wang$^{2}$ and Zhaojun Wang$^{1}$ \\
$^{1}$School of Statistics and Data Science, KLMDASR, LEBPS, and LPMC,\\ Nankai University\\
$^{2}$Department of Statistics, Nanjing Audit University
}
\date{}

\begin{document}
\maketitle

\begin{abstract}
In this paper, we investigate alpha testing for high-dimensional linear factor pricing models. We propose a spatial sign-based max-type test to handle sparse alternative cases. Additionally, we prove that this test is asymptotically independent of the spatial-sign-based sum-type test proposed by \cite{liu2023high}. Based on this result, we introduce a Cauchy Combination test procedure that combines both the max-type and sum-type tests. Simulation studies and real data applications demonstrate that the new proposed test procedure is robust not only for heavy-tailed distributions but also for the sparsity of the alternative hypothesis.
\end{abstract}
\section{Introduction}

In the realm of investment analysis and finance, linear factor pricing models serve as fundamental tools for understanding and predicting asset returns. The Capital Asset Pricing Model (CAPM), formulated by \cite{Sharpe1964}, pioneered this field by postulating that a stock's return is primarily driven by its exposure to market risk, quantified by beta ($\beta$). This single-factor approach laid the groundwork for modern portfolio theories. However, as financial markets evolved, researchers recognized the need for a more comprehensive framework. Stephen Ross's Arbitrage Pricing Theory (APT) \citep{ross1976} introduced the concept of multiple factors, suggesting that asset returns are influenced by a diverse set of macroeconomic variables. This multi-factor approach has gained traction, enhancing predictive accuracy and capturing market nuances.


In linear factor pricing models, the intercept (alpha) represents unexplained asset returns, capturing average residuals or mean variation not captured by factors. It reflects idiosyncratic characteristics, independent of market fluctuations. The intercept's significance varies across models and contexts. Researchers assess if alpha differs significantly from zero to evaluate model adequacy. A significant alpha hints at model limitations requiring adjustment. Conversely, an insignificant alpha indicates the model adequately explains returns, possibly sufficient for investment analysis. This evaluation ensures factor pricing models' relevance and applicability in diverse empirical settings.

In the context of financial markets, where the number of securities ($N$) frequently surpasses the length of the observation period ($T$), traditional multivariate F-tests based on joint normality assumptions, pioneered by \citet{GIBBONS1989A} and extended by others, become inadequate. Recognizing this limitation, recent research has focused on developing tests tailored for high-dimensional settings.
\citet{pesaran2023testing} contributed by introducing a sum-type test statistic that circumvents the need for a fixed N by substituting the sample covariance matrix with an identity matrix, demonstrating robustness under dense alternatives. \citet{lan2018testing} proposed a random projection method that accommodates non-sparse covariance matrices of idiosyncratic terms.
For sparse alternatives, \citet{Gungor2013TestingLF} and \citet{feng2022high} advocated max-type test statistics, while \citet{Fan2015} enhanced test power through a specific power enhancement procedure. \citet{Yu2023PE} innovated further by leveraging the thresholding covariance estimator of \citet{Fan2011LargeCE} to devise a novel Wald-type test and a Cauchy combination test blending this with a max-type test. Lastly, \citet{xia2023adaptive} explored the use of L-statistics as an approach to tackle sparse alternatives. These advancements reflect the evolving landscape of statistical testing in high-dimensional financial markets.

All the aforementioned high-dimensional test procedures hinge on the assumption that the error term is normally distributed or adheres to independent component models. As is widely known, these models are confined to light-tailed distributions, excluding many renowned heavy-tailed distributions such as the multivariate t-distribution and multivariate mixture normal distribution. Unfortunately, financial data often exhibit heavy tails, making it imperative to develop efficient test procedures that perform well under heavy-tailed distributions. To capture the heavy-tail characteristic of security returns, a popular distribution family in the literature is the elliptical distribution family. Indeed, \citet{Chamberlain1983} demonstrated that a mean-variance analysis of the CAPM aligns with investors' portfolio decision-making if and only if the returns are elliptically distributed. Furthermore, in the case of elliptical returns, the CAPM remains theoretically valid.

For elliptical distributions, classic spatial-sign-based procedures have proven to be highly robust and efficient in traditional multivariate analysis, as overviewed by \citet{Oja2010Multivariate}. Recent literature has also shown that these spatial-sign-based procedures excel in high-dimensional settings. Specifically, \citet{wang2015high}, \citet{feng2016}, and \citet{f2021} have proposed spatial-sign-based test procedures for the high-dimensional one-sample location problem. Additionally, \citet{Feng2016Multivariate} and \citet{h2022} have addressed the high-dimensional two-sample location problem using spatial-sign-based methods. Moreover, \citet{zou2014multivariate}, \citet{fl2017} and \cite{zhang2022robust} extended the spatial-sign-based method to the high-dimensional sphericity test, while \citet{paindaveine2016high} and \cite{zhao2023spatial} considered high-dimensional white noise tests.

For testing alpha in high-dimensional linear factor pricing models, \cite{liu2023high} and \cite{zhao2022high} proposed two spatial-sign-based test procedures that also exhibit good performance for heavy-tailed distributions. Additionally, \cite{zhao2023} extended \cite{liu2023high}'s test procedure to time-varying factor models. However, these three tests are all sum-type procedures, which perform well under dense alternatives but have less power under sparse alternatives. To address this issue, we aim to construct a spatial-sign-based max-type test procedure. For high-dimensional location test problems, \cite{cheng2023} established a Gaussian approximation for the sample spatial median over the class of hyperrectangles and constructed a max-type test procedure using a multiplier bootstrap algorithm. However, their proposed test statistic is not scalar-invariant, and they did not provide the limiting null distribution. Furthermore, the multiplier bootstrap algorithm is time-consuming. To overcome these limitations, \cite{liu2024spatial} constructed a new spatial-sign-based max-type test statistic that is scalar-invariant and has a simple limiting null distribution. Therefore, we extend the method in \cite{liu2024spatial} to the alpha testing problem in linear factor models. We establish the limiting null distribution of the newly proposed max-type test procedure and demonstrate its consistency for sparse alternatives.

Furthermore, to construct a doubly robust test procedure that is robust not only to heavy-tailed distributions but also to the sparsity levels of the alternatives, we demonstrate the asymptotic independence between the sum-type spatial-sign-based test \citep{liu2023high} and our newly proposed max-type spatial-sign-based test. Based on this, we construct a Cauchy combination test procedure. Simulation studies and a real data application both demonstrate the robustness of our proposed Cauchy combination test procedure.

The main contributions of this paper are threefold:
\begin{itemize}
\item[1.] We establish theoretical results under a broader model that encompasses not only elliptical distributions but also independent component models, broadening the applicability of our methods.
\item[2.] We develop a spatial-sign-based max-type test procedure that is robust against heavy-tailed distributions and performs well under sparse alternatives. The theoretical findings are not insignificant. Contrary to the typical mean testing problem, we must address the factors that necessitate rescaling the test statistic, a task that considerably increases the complexity of the proofs.
\item[3.] We demonstrate the asymptotic independence between the spatial-sign-based sum-type and max-type test procedures under both the null and alternative hypotheses. Our newly proposed Cauchy Combination test procedure exhibits double robustness against heavy-tailed distributions and the sparsity level of alternatives.
\end{itemize}

The remainder of the paper is structured as follows. Section 2 provides a concise overview of recent significant alpha test procedures and introduces a spatial-sign-based max-type test procedure. In Section 3, we demonstrate the asymptotic independence between the sum-type spatial-sign-based test and our newly proposed max-type spatial-sign-based test, and subsequently construct a Cauchy combination test. Section 4 presents simulation studies, while Section 5 examines a real data application. We conclude this
paper in Section 6. All proofs are included in the Appendix.

\textsc{Notations.} For $k$-dimensional vector $\boldsymbol{a}$, we use the notation $\|\boldsymbol{a}\|$ and $\|\boldsymbol{a}\|_{\infty}$ to denote its Euclidean norm and maximum-norm respectively. For a $m \times n$ matrix $\mathbf{M}=\left(m_{j \ell}\right)_{m \times n}$, the 1- and 2-norms of $\mathbf{M}$ are $\|\mathbf{M}\|_1=\max _{1 \leqslant \ell \leqslant n} \sum_{j=1}^{m}\left|m_{j \ell}\right|$ and $\|\mathbf{M}\|_2=\left\{\lambda_{\max }\left(\mathbf{M}^{\top} \mathbf{M}\right)\right\}^{1 / 2}$. The Frobenius norm of $\mathbf{M}$ is $\|\mathbf{M}\|_F=\left\{\sum_{j=1}^{m} \sum_{\ell=1}^{n} m_{j \ell}^2\right\}^{1 / 2}$.
Denote $a_n \lesssim b_n$ if there exists constant $C, a_n \leq C b_n$ and $a_n \asymp b_n$ if both $a_n \lesssim b_n$ and $b_n \lesssim a_n$ hold. For two sequences of numbers $\left\{a_n \geq 0 ; n \geq 1\right\}$ and $\left\{b_n>0 ; n \geq 1\right\}$, we write $a_n \ll b_n$ if $\lim _{n \rightarrow \infty} \frac{a_n}{b_n}=0$. The following assumption will be imposed. Let $\psi_\alpha(x)=\exp \left(x^\alpha\right)-1$ be a function defined on $[0, \infty)$ for $\alpha>0$. Then the Orlicz norm $\|\cdot\|_{\psi_\alpha}$ of a $\boldsymbol{X}$ is defined as $\|\boldsymbol{X}\|_{\psi_\alpha}=\inf \left\{t>0, \mathbb{E}\left\{\psi_\alpha(|\boldsymbol{X}| / t)\right\} \leqslant 1\right\}$. Let $\operatorname{tr}(\cdot)$ be a trace for matrix, $\lambda_{\text {min }}(\cdot)$ and $\lambda_{\max }(\cdot)$ be the minimum and maximum eigenvalue for symmetric martix. $\mathbf{I}_k$ represents a $k$-dimensional identity matrix, and $\operatorname{diag}\left\{v_1, v_2, \cdots, v_k\right\}$ represents the diagonal matrix with entries $\boldsymbol{v}=\left(v_1, v_2, \cdots, v_k\right)$. For $a, b \in \mathbb{R}$, we write $a \wedge b=\min \{a, b\}$.
\section{Spatial-sign based Max-type test procedure}
For these $N$ assets, we further write the form of the linear asset pricing model in the following multivariate linear regression \citep{GIBBONS1989A}
\begin{align}\label{mod}
Y_{it}=\alpha_i+\bm\beta_i^\top \f_t+\varepsilon_{it}, i=1,\cdots,N, t=1,\cdots,T.
\end{align}
The intercept term $\alpha_i$ captures the excess return of the $i-$th security.
We want to test whether the observed traded factors are sufficient to price all assets, i.e. testing
\begin{align} \label{h1}
H_0: \boldsymbol  \alpha=\mathbf{0}~~ \text{versus}~~ H_1: \boldsymbol  \alpha\not=\mathbf{0}.
\end{align}
For notational convenience, we write $\Y_{i\cdot}=(Y_{i1}, \cdots, Y_{iT})^\top\in\mR^T$,
$\Y_{\cdot t}=(Y_{1t}, \cdots, Y_{Nt})^\top\in\mR^N$ and $\Y=(\Y_{1\cdot}, \cdots, \Y_{N\cdot})\in\mR^{T\times N}$. And $\bmv_{i\cdot}=(\varepsilon_{i1}, \cdots, \varepsilon_{iT})^\top\in\mR^T$,
$\bmv_{\cdot t}=(\varepsilon_{1t}, \cdots, \varepsilon_{Nt})^\top\in\mR^N$ and $\bmv=(\bmv_{1\cdot}, \cdots, \bmv_{N\cdot})\in\mR^{T\times N}$. Here $\bmv_{\cdot t}$ has mean $\mathbf{0}$ and scatter matrix $\L$.
In addition, let $\alpha=(\alpha_1, \cdots, \alpha_N)^\top\in\mR^N$, which
collects all the intercepts for every asset, and let $\X=(\f_1, \cdots, \f_T)^\top\in\mR^{T\times p}$ be the common factor matrix.
When the number of assets $N$ is fixed, the traditional GRS test statistic \citep{GIBBONS1989A} is
\begin{align*}
GRS=\frac{T-N-p}{N}\left(\frac{\boldsymbol  1^\top_T\P_{\X}\boldsymbol  1_T}{T}\right)\hat{\boldsymbol  \alpha}^\top \hat{\L}^{-1}\hat{\bma}
\end{align*}
where $\boldsymbol  1_T=(1,\cdots,1)^\top\in\mathcal{R}^{T}$, $\P_{\X}=\I_T-\X(\X^\top\X)^{-1}\X^\top$ and
\begin{align*}
\hat{\alpha}_i=\Y_{i\cdot}^\top \left(\frac{\P_{\X}\boldsymbol  1_T}{\boldsymbol  1_T^\top \P_{\X}\boldsymbol  1_T}\right), \hat{\bmv}_{i\cdot}=\P_{\X}(\Y_{i \cdot}-\hat{\alpha}_i \boldsymbol  1_T), \hat{\L}=\frac{1}{T}\sum_{t=1}^T\hat{\bmv}_{\cdot t}\hat{\bmv}_{\cdot t}^\top.
\end{align*}
Under the normality assumption of $\bmv_{\cdot t}$, the GRS test statistic is distributed exactly as $F(T-N-p,N)$ under the null hypothesis. However, the traditional GRS test can not be applied when $N>T$ because $\hat{\L}$ is not invertible. To this end, \citet{pesaran2023testing} proposed a new test (hereafter, the PY test)
\begin{align*}
T_{PY}=\frac{N^{-1/2}(\sum_{i=1}^N t_i^2-\frac{v}{v-2})}{\frac{v}{v-2}\sqrt{\frac{2(v-1)}{v-4}(1+(N-1)\tilde{\rho}_{N,T}^2)}},t_i^2=\frac{\hat{\alpha}_i^2(\bm 1_T^\top \P_{\X}\bm
  1_T)}{v^{-1}\hat{\bmv}_{i\cdot}^\top\hat{\bmv}_{i\cdot}}, v=T-p-1
\end{align*}
 by replacing $\hat{\L}$ with the threshold covariance estimator of \citet{fan2011high} and $\tilde{\rho}_{N,T}$ is the corresponding correlation estimator. However, the PY test performs not very well for heavy-tailed distributions. So \citet{liu2023high} proposed  the following spatial sign-based test statistic:
\be\label{eq:TS}
T_{SS}=\frac{Q-\delta_Q}{\sqrt{2\wh{\tr(\R^2)}}},
\ee
where
\be\label{eq:Q}
Q=N(\h\trans\h)^{-1}\sum_{t_1, t_2=1, t_1\ne t_2}^T
h_{t_1}h_{t_2}\wh{\U}_{t_1}^\top\wh{\U}_{t_2},
\ee
and for any $t=1, \dots, T$,
$h_t$ is the $t$th element of $\h=\M_{\X}\1_T$,
$\wh{\U}_t=U\{\wh\D^{-1/2}(\Y_t-\wh\B\f_t)\}$. where $\hat{\U}_i=U(\hat{\D}^{-1/2}\Z_i)$, $\Z=\P_{\X}\Y=(\Z_{1},\cdots,\Z_{T})^{\top}$, $\Z_{t}=(Z_{1t},\cdots,Z_{Nt})^{\top}=\Y_{\cdot t}- \hat{\boldsymbol \beta} \f_t$. Here  $\hat{\boldsymbol  \beta}$ is  the least-square estimator, i.e.
\begin{align*}
\hat{\boldsymbol  \beta}=(\hat{\boldsymbol  \beta}_1,\cdots,\hat{\boldsymbol  \beta}_N)^\top, ~~\hat{\boldsymbol  \beta}_i=(\X^\top\X)^{-1}\X^\top\Y_{i\cdot}, i=1,\cdots,N.
\end{align*}
And $\hat{\D}$ is the estimator of $\D$ with the sample $\{\hat{\bmv}_{\cdot t}\}_{t=1}^T$ by the following algorithm:
\begin{itemize}
\item[(i)] $\boldsymbol  \xi_t \leftarrow \D^{-1/2}\hat{\bmv}_{\cdot t}$,
~~$t=1,\cdots,T$;
\item[(ii)] $\D \leftarrow N
\D^{1/2}\diag\{T^{-1}\sum_{t=1}^{T}U(\boldsymbol  \xi_t)U(\boldsymbol  \xi_t)^\top \}\D^{1/2}$.
\end{itemize}
Here,
$\R=\D^{-1/2}\bSig\D^{-1/2}$ is the correlation matrix, and
\bse
\wh{\tr(\R^2)}\equiv
\frac{N^2}{\h\trans\h(\h\trans\h-1)}\sum_{t_1, t_2=1, t_1\ne t_2}^T h^2_{t_1}h^2_{t_2}
\left\{U(\wh\D^{-1/2}
\weps_{t_1}^{(t_1,t_2)})\trans
U(\wh\D^{-1/2}\weps_{t_2}^{(t_1,t_2)})\right\}^2,
\ese
where $\weps_{t_1}^{(t_1,t_2)}=\Y_{t_1}-\wh\B_{t_1}^{(t_1t_2)}\f_{t_1}$,
$\weps_{t_2}^{(t_1,t_2)}=\Y_{t_2}-\wh\B_{t_2}^{(t_1t_2)}\f_{t_2}$, and $\wh\B_{t_1}^{(t_1t_2)}$
and $\wh\B_{t_2}^{(t_1t_2)}$ are the least-square estimators of $\B$ based
on the first half and second half of the sample
$\{(\Y_t,\f_t)\}_{t\not\equiv t_1,t_2}$, respectively.

Both PY and SS test are sum-type test procedures, which has good performance under dense alternatives,i.e. the number of securities with nonzero alpha are very large. However, they has less power when the alternative is sparse, i.e. only a few securities has nonzero alpha. So \cite{feng2022high} proposed the following max-type test statistic
\begin{align*}
T_{MAX}=\max_{1\le i \le N} t_i^2.
\end{align*}
They established its asymptotic null distribution and showed that it has very good performance under sparse alternatives. However, $T_{MAX}$ do not performs very well for heavy-tailed distributions. So we would proposed a spatial-sign based max-type test procedure in this paper.

Note that $T^{-1}\omega_T\hat{\bm \alpha}$ is the sample mean of $\{\Z_t\}_{t=1}^T$ where $\omega_T=1_T^\top \P_{\X}\bm
  1_T$. As known to all, the sample mean has good performance for light-tailed distributions, but is not very robust for heavy-tailed distributions. In contrast, the spatial median has very good performance for heavy-tailed distributions \citep{Oja2010Multivariate}.

So similar to \citet{Feng2016Multivariate}, we first estimate the spatial median and diagonal matrix $\D$ with the sample  $\{\Z_t\}_{t=1}^T$ by solving the following equations:
\begin{align}
&\frac{1}{T}\sum_{t=1}^T U(\D^{-1/2}(\Z_t-\bm \theta))=0\\
&\frac{1}{T}\sum_{t=1}^T\diag\{ U(\D^{-1/2}(\Z_t-\bm \theta))U(\D^{-1/2}(\Z_t-\bm \theta))^\top\}=\frac{1}{N}\I_N
\end{align}
Similary, we adopt the following algorithm to solve the above equations:
\begin{itemize}
\item[(i)] $\boldsymbol  \xi_t \leftarrow \D^{-1/2}(\Z_{t}-\bm \theta)$,
~~$t=1,\cdots,T$;
\item[(ii)] $\bm \theta \leftarrow \bm \theta+\frac{\D^{1/2}\sum_{t=1}^T U(\bm\xi_t)}{\sum_{t=1}^T ||\bm\xi_t||^{-1}}$
\item[(iii)] $\D \leftarrow N
\D^{1/2}\diag\{T^{-1}\sum_{t=1}^{T}U(\boldsymbol  \xi_t)U(\boldsymbol  \xi_t)^\top \}\D^{1/2}$.
\end{itemize}
Denote the result estimator as $\hat{\bm \theta}$ and $\hat{\D}$. Next, we will establish the theoretial results of $\hat{\bm \theta}$.

{ Let $\mathcal{F}_{-\infty}^0$ and $\mathcal{F}_k^{\infty}$ denote the $\sigma$-algebras generated by $\big\{\f_t : -\infty \leq t \leq 0\big\}$ and $\big\{\f_t: k \leq t \leq \infty\big\}$, respectively. Let $\alpha(k) \equiv \sup _{A \in \mathcal{F}_{-\infty}^0, B \in \mathcal{F}_k^{\infty}}|P(A) P(B)-P(A B)|$ denote the $\alpha$-mixing coefficient. We need the following conditions:}
\begin{itemize}
\item[(C1)] The $p$-dimensional vector of common factors $\f_t$ is strictly stationary and
  distributed independently of the errors $\varepsilon_{it'}$,
for all  $i=1,\cdots,N$ and all $t,t'=1,\cdots,T$. The
number of factors $p$ is fixed
and $\f_t\trans \f_t\le K<+\infty$, for a constant $K$ and all
$t=1,\cdots,T$.
The matrix $T^{-1}(\1_T, \F)\trans (\1_T, \F)$
is positive definite, and as $T\to \infty$,
$T^{-1}\1_T\trans\M_\F\1_T>\tau_{\min}$ for some positive constant
$\tau_{\min}$. { There exist $c_2,c_3 >0$, such
that for any positive integer $k$, the $\alpha$-mixing coefficient
$\alpha(k) \leq  \exp(-c_2k^{c_3})$. }
\item[(C2)] We consider the following model for error term:
\begin{align}\label{modelx}
\bmv_{\cdot t}=v_t\mathbf\Gamma \boldsymbol W_t,
\end{align}
where $\boldsymbol W_t$ is a
p-dimensional random vector with independent components, $\mathbb{E}(\boldsymbol W_t)=0$, $\mathbf \Sigma=\mathbf \Gamma\mathbf\Gamma^\top$, $v_t$ is a nonnegative univariate random variable and is independent with the spatial sign of $\boldsymbol W_t$ and { $0<E(v_t^2)<\infty$.}
\item[(C3)] (i) $W_{t, 1}, \ldots, W_{t, N}$ are i.i.d. symmetric random variables with $\mathbb{E}\left(W_{t, j}\right)=0, \mathbb{E}\left(W_{t, j}^2\right)=$ 1 , and $\left\|W_{t, j}\right\|_{\psi_\alpha} \leqslant c_0$ with some constant $c_0>0$ and $1 \leqslant \alpha \leqslant 2$. (ii) Let $\D$ is the diagonal matrix of $\bms$ and $r_t=||\D^{-1/2}\bmv_{\cdot t}||$. The moments $\zeta_k=\mathbb{E}\left(r_t^{-k}\right)$ for $k=1,2,3,4$ exist for large enough $N$. In addition, there exist two positive constants $\underline{b}$ and $\bar{B}$ such that $\underline{b} \leqslant \lim \sup_N \mathbb{E}\left(r_t / \sqrt{N}\right)^{-k} \leqslant \bar{B}$ for $k=1,2,3,4$.
\item[(C4)] (i) The shape matrix $\R=\mathbf D^{-1/2}{\bf \Gamma\Gamma}^\top \mathbf D^{-1/2}=\left(\sigma_{j \ell}\right)_{N \times N}$  satisfies $\max _{j=1,\cdots,N}\sum_{\ell=1}^N\left|\sigma_{j \ell}\right| \leqslant a_0(N)$. We assume that $a_0(N)\asymp N^{1-\delta}$, ${\color{black}0<\delta\leq1/2}$,  $\log N=o(T^{1/5})$ and $\log T=o(N^{1/3 \wedge \delta})$. In addition, $\lim\inf_{N\rightarrow \infty}\min_{j=1,2,\cdots,N}{\color{black}d}_j>\underline{d}$ for some constant $\underline d>0$, where $\mathbf D=\operatorname{diag}\{d_1^2,d_2^2,\cdots,d_N^2\}$. (ii)   For some $\varrho \in(0,1)$, assume $|\sigma_{ij}|\leq \varrho$ for all $1\leq i<j\leq N$ and $N
    \geq 2$. Suppose $\left\{\delta_N ; N \geq 1\right\}$ and $\left\{\kappa_N ; N \geq 1\right\}$ are positive constants with $\delta_N=o(1 / \log N)$ and $\kappa=\kappa_N \rightarrow 0$ as $N \rightarrow \infty$. For $1 \leq i \leq N$, define $B_{N, i}=\left\{1 \leq j \leq N ;\left|\sigma_{i j}\right| \geq \delta_N\right\}$ and $C_N=\left\{1 \leq i \leq N ;\left|B_{N, i}\right| \geq N^\kappa\right\}$. We assume that $\left|C_N\right| / N \rightarrow 0$ as $N \rightarrow \infty$.
\end{itemize}

\begin{remark}{}
Condition (C1) is identical to Assumption (A1) in \cite{liu2023high}, which is extensively used in high-dimensional alpha testing procedures, such as those found in \cite{Fan2015}, \cite{lan2018testing}, and \cite{pesaran2023testing}. It necessitates independence between the factors and the errors, as well as uniform boundedness of $\f_t^\top\f_t$. The error model (\ref{modelx}) outlined in Condition (C2) corresponds to model (3) in \cite{cheng2023}, encompassing a broad spectrum of commonly employed multivariate models and distribution families, including the independent components model \citep{nordhausen2009signed,ilmonen2011semiparametrically,yao2015sample} and the family of elliptical distributions \citep{hallin2006semiparametrically,Oja2010Multivariate,fang2018symmetric}. Condition (C3) aligns with Conditions (C.1) and (C.2) in \cite{cheng2023}. Here, the symmetric assumption ensures that the spatial median of $\{\Z_t\}_{t=1}^T$ is $T^{-1}\omega_T \bm \alpha$. Condition (C3)-(ii) generalizes Assumption 1 in \cite{zou2014multivariate}, indicating that $\zeta_k \asymp N^{-k/2}$. Further elaboration on Condition (C3) can be found in \cite{cheng2023}. Condition (C4) stipulates that the correlation among these variables cannot be excessively high. It is straightforward to verify that both banded matrices with a fixed band and matrices with an AR(1) structure satisfy this condition.
\end{remark}

\begin{theorem}\label{th1}
Under the null hypothesis and Conditions (C1)-(C4), we have
\begin{align*}
P\left(T||\hat{\D}^{-1/2}\hat{\bm \theta}||^2_\infty\zeta-2\log N+\log\log N\le x\right)\to \exp \left\{-\frac{1}{\sqrt{\pi}}e^{-x/2}\right\}
\end{align*}
where { $\zeta=N\{E(r_t^{-1})\}^2/\eta_{\omega}$, $\eta_{\omega}=1-2\eta E(r_t^{-1})E(r_t)+\eta E(r_s^{-2})E(r_t^2)$ and $T^{-1}\omega_{T}\cp \omega=1-\eta$.}
\end{theorem}
{ 
We estimate $\zeta$ by  $$\hat{\zeta}=\frac{N\{\widehat{E(r_t^{-1})}\}^2}{1-2T^{-1}\1_{T}^{\top}\F(\F^{\top}\F)^{-1}\F^{\top}\1_T\widehat{E(r_t^{-1})}\widehat{E(r_t)}+T^{-1}\1_{T}^{\top}\F(\F^{\top}\F)^{-1}\F^{\top}\1_T\widehat{E(r_t^2)}\widehat{E(r_t^{-2})}}$$ and
\begin{align*}
\widehat{E(r_t^2)}=&T^{-1}\sum_{t=1}^T ||\tilde{\bmv}_{t\cdot}||^2,\,\,\,
&\widehat{E(r_t^{-1})}=T^{-1}\sum_{t=1}^T ||\tilde{\bmv}_{t\cdot}||^{-1},\\
\widehat{E(r_t)}=&T^{-1}\sum_{t=1}^T ||\tilde{\bmv}_{t\cdot}||,\,\,\,
&\widehat{E(r_t^{-2})}=T^{-1}\sum_{t=1}^T ||\tilde{\bmv}_{t\cdot}||^{-2}
\end{align*}
Here $\tilde{\bmv}_{t\cdot}=\hat\D^{-1/2}(\Z_t-\hat{\bm \theta})$.}

\begin{prop}\label{pro1}
Under Conditions (C1)-(C4), we have $\hat{\zeta}/\zeta\cp 1$.
\end{prop}

So we proposed the following spatial-sign based max-type test statistic
\begin{align}\label{mt}
T_{SM}=T||\hat{\D}^{-1/2}\hat{\bm \theta}||^2_\infty\hat\zeta-2\log N+\log\log N
\end{align}
According to Theorem \ref{th1} and Proposition \ref{pro1}, we have
\begin{align*}
P\left(T_{SM}\le x\right)\to \exp \left\{-\frac{1}{\sqrt{\pi}}e^{-x/2}\right\}\doteq G(x).
\end{align*}
So, the $p$-value of the SM test is
\begin{align}
p_{SM}=1-G(T_{SM})
\end{align}
and we reject the null hypothesis when $p_{SM}\le \gamma$ at significant level $\gamma$.

\begin{theorem}
Under Conditions (C1)-(C4), for some sufficiently large constant $C>0$,
if $\|\bm \alpha\|_{\infty} \geq C\sqrt{\log N/T}${,  $\|\bm \alpha\|^2 =O (NT^{-1}\log N)$ and $\lambda_{\max}(\mathbf{R})=o(N/(\log N)^2)$}, we have the power of $T_{SM}$ goes to one.
\end{theorem}
\section{Robust Cauchy combination test}
In practice, we cannot determine whether the alternative hypothesis is dense or sparse. Therefore, it is necessary to combine the sum-type test and the max-type test to construct a novel test that is robust to the sparsity of the alternative hypothesis. \cite{feng2022high} proposed a test (hereinafter referred to as COM) by considering the minimum $p$-values of the MAX and PY tests to be smaller than $\gamma/2$. However, both the MAX and PY tests lack robustness when dealing with heavy-tailed distributions. As a result, we need to develop a new test procedure that is not only robust to the sparsity of the alternative hypothesis but also robust to heavy-tailed distributions. Consequently, we first analyze the relationship between the spatial-sign-based sum-type test (\ref{eq:TS}) and the max-type test (\ref{mt}).

 We impose the following condition:
\begin{itemize}
	{ \item [(C5)] $\operatorname{tr}\left(\mathbf{R}^4\right)=o\left\{\operatorname{tr}^2\left(\mathbf{R}^2\right)\right\}$; $T^{-2} N^2 / \operatorname{tr}\left(\mathbf{R}^2\right)=O(1)$; $\operatorname{tr}\left(\mathbf{R}^2\right)-N=o\left(T^{-1} N^2\right)$.}
	\item [(C6)] There exist $C>0$ so that $\max _{1 \leq i \leq N} \sum_{j=1}^N \sigma_{i j}^2 \leq$ $(\log N)^C$ for all $N \geq 3 ; N^{-1 / 2}(\log N)^C \ll \lambda_{\min }(\mathbf{R}) \leq \lambda_{\max }(\mathbf{R}) \ll \sqrt{N}(\log N)^{-1}$ and $\lambda_{\max }(\mathbf{R}) / \lambda_{\min }(\mathbf{R})=$ $O\left(N^\tau\right)$ for some $\tau \in(0,1 / 4)$.
\end{itemize}
	{ Note that Condition (C5) is identical to Assumption (A3) in \cite{liu2023high}, while Condition (C6) is identical to Assumption (2.3) in \cite{feng2022asymptotic}.}
\begin{theorem}\label{thm3}
Under Conditions (C1)-(C6) and $\log N=o({ T^{1/10}})$, we have $T_{SS}$ is asymptotically independent with $T_{SM}$ under the null hypothesis, i.e.
\begin{align*}
P\left(T_{SS}\le x, T_{SM}\le y\right)\to \Phi(x)G(y).
\end{align*}
\end{theorem}
According to Theorem \ref{thm3}, we suggest combining the corresponding $p$-values by using truncated Cauchy Combination Method \citep{liu2020}, to wit,
\begin{align*}
    p_{CC}&=1-F[0.5\tan\{(0.5-p_{SS})\pi\}I(p_{SS}<0.5)+0.5\tan\{(0.5-p_{SM})\pi I(p_{SM}<0.5)\}]
\end{align*}
where $F(\cdot)$ is the CDF of the the standard Cauchy distribution. If the final $p$-value is less than some pre-specified significant level $\gamma\in(0,1)$, then we reject $H_0$. We called this test procedure CC test hereafter.

To analysis the power performance of the new proposed CC test, we also demonstrate that $T_{SS}$ are also asymptotically independent with $T_{SM}$ under some special alternatives.
{\color{black}We consider the relationship between $T_{S S}$ and $T_{SM }$ under local alternative hypothesis:
\begin{align}\label{H_1_comb}
H_1:\|\boldsymbol{\alpha}\|^2=O\left(\zeta_1^{-2} N^{-1} T^{-1} \sigma_{SS}\right),\left\|\mathbf{R}^{1 / 2} \boldsymbol{\alpha}\right\|^2=o\left(\zeta_1^{-2} N^{-1} T^{-1} \sigma_{SS}^2\right)\n\\
 \text { and }|\mathcal{A}|=o\left(\frac{\lambda_{\min }(\mathbf{R})\left[\operatorname{tr}\left(\mathbf{R}^2\right)\right]^{1 / 2}}{(\log N)^C}\right),
\end{align}
where $\mathcal{A}=\left\{i \mid \alpha_i \neq 0,1 \leq i \leq N\right\},$ $\boldsymbol{\alpha}=\left(\alpha_1, \alpha_2, \cdots, \alpha_N\right)^{\top}$ and $\sigma_{SS}=\sqrt{2\tr(\R^2)}$. The following theorem establish the asymptotic independence between $T_{S M}$ and $T_{SS}$ under this special alternative hypothesis.}
\begin{theorem}\label{thm4}
Under Conditions (C1)-(C6), $\log N=o({ T^{1/10}})$, and the alternative hypothesis \eqref{H_1_comb}, we have
\begin{align*}
P\left(T_{SS}\le x, T_{SM}\le y\right)\to P\left(T_{SS}\le x\right)P\left(T_{SM}\le y\right)
\end{align*}
\end{theorem}

According to \cite{li2023}, the Cauchy combination-based test has more power than the test based on the minimum of $p_{SM}$ and $p_{SS}$, which is also known as the minimal p-value combination. This is represented as $\beta_{M\wedge S, \alpha}=P(\min\{{\rm p}_{SM},{\rm p}_{SS}\}\leq 1-\sqrt{1-\alpha})$.

It is clear that:
\begin{align}\label{power_H1}
\beta_{M\wedge S, \alpha} &\ge P(\min\{{\rm p}_{SM},{\rm p}_{SS}\}\leq \alpha/2)\nonumber\\
&= \beta_{SM,\alpha/2}+\beta_{SS,\alpha/2}-P({\rm p}_{SM}\leq \alpha/2, {\rm p}_{SS}\leq \alpha/2)\nonumber\\
&\ge \max\{\beta_{SM,\alpha/2},\beta_{SS,\alpha/2}\}.
\end{align}

On the other hand, under the local alternative hypothesis (\ref{H_1_comb}), we have:
\begin{align}\label{power_H1np}
\beta_{M\wedge S, \alpha} \ge \beta_{SM,\alpha/2}+\beta_{SS,\alpha/2}-\beta_{SM,\alpha/2}\beta_{SS,\alpha/2}+o(1),
\end{align}
which is due to the asymptotic independence implied by Theorem \ref{thm4}.

For a small $\alpha$, the difference between $\beta_{SM,\alpha}$ and $\beta_{SM,\alpha/2}$ is small, and the same applies to $\beta_{SS,\alpha}$. Therefore, according to equations \eqref{power_H1} and \eqref{power_H1np}, the power of the adaptive test is at least as large as, or even significantly larger than, that of either the max-type or sum-type test. For a detailed comparison of the performance of each test type under varying conditions of sparsity and signal strength, please refer to Table 1 in \cite{ma2024testing}.

\section{Simulation}
This experiment aims to replicate the well-known Fama-French three-factor model, with a focus on incorporating strong serial correlation and heterogeneous variance into the factors $f_t$. To achieve this, we adapt and modify an example studied in Section 5.1 of \cite{pesaran2023testing}.

The response variables $Y_{it}$ are generated using the Linear Factor Pricing Model (LFPM) outlined in (\ref{mod}), specifically with $p=3$ factors:
$$
Y_{it} = \alpha_i + \sum_{k=1}^3 \beta_{ik} f_{kt} + \varepsilon_{it},
$$
where $f_{t1}$, $f_{t2}$, and $f_{t3}$ represent the Fama-French three factors: Market factor, SMB (Size factor), and HML (Value factor). A detailed explanation of these factors is provided in Section \ref{Emp}.

To generate the factors, we employ an autoregressive process with conditional heteroskedasticity, utilizing the GARCH(1,1) model. The coefficients used are consistent with those in \cite{pesaran2023testing}. Specifically, the factors are generated as follows:
\begin{align*}
f_{1t} &= 0.53 + 0.06 f_{1,t-1} + h_{1t}^{1/2} \zeta_{1t}, & \text{(Market factor)} \\
f_{2t} &= 0.19 + 0.19 f_{2,t-1} + h_{2t}^{1/2} \zeta_{2t}, & \text{(SMB factor)} \\
f_{3t} &= 0.19 + 0.05 f_{3,t-1} + h_{3t}^{1/2} \zeta_{3t}, & \text{(HML factor)}
\end{align*}
where $\zeta_{kt}$ are drawn from a standard normal distribution. The variance terms $h_{kt}$ are generated using:
\begin{align*}
h_{1t} &= 0.89 + 0.85 h_{1,t-1} + 0.11 \zeta_{1,t-1}^2, & \text{(Market factor)} \\
h_{2t} &= 0.62 + 0.74 h_{2,t-1} + 0.19 \zeta_{2,t-1}^2, & \text{(SMB factor)} \\
h_{3t} &= 0.80 + 0.76 h_{3,t-1} + 0.15 \zeta_{3,t-1}^2, & \text{(HML factor)}
\end{align*}
Consistent with the approach in \cite{pesaran2023testing}, we simulate the above process over the periods $t = -49, \ldots, 0, 1, \ldots, T$, initializing with $f_{k,-50} = 0$ and $h_{k,-50} = 1$ for $k = 1, 2, 3$. For our final experiments, we utilize the simulated data from observations $t = 1, \ldots, T$.

The errors are generated from four scenarios with $\bms=(0.5^{|i-j|})_{1\le i,j\le N}$:
\begin{itemize}
\item[(I)] Multivariate normal distribution. $\bm \varepsilon_{\cdot t}\sim N(\bm\theta,\bms)$.
\item[(II)] Multivariate $t$-distribution $t_{N,3}$.   $\bm \varepsilon_{\cdot t}$'s are generated from standardized $t_{N,3}/\sqrt{3}$ with mean zero and scatter matrix $\bms$.
\item[(III)] Multivariate mixture normal distribution $\mbox{MN}_{N,\kappa,9}$. $\bm \varepsilon_{\cdot t}$'s are generated from standardized  $[\kappa
N(\bm 0,\bms)+(1-\kappa)N(\bm 0,9\bms)]/\sqrt{\kappa+9(1-\kappa)}$, denoted
by $\mbox{MN}_{N,\gamma,9}$. $\kappa$ is chosen to be 0.8.
\item[(IV)] Independent Component Model. $\bm \varepsilon_{\cdot t}=\bms^{1/2}\bm\epsilon_{\cdot t}$, $\bm\epsilon_{\cdot t}=(\epsilon_{t1},\cdots,\epsilon_{tN})^\top$ where $\epsilon_{ti}, i=1,\cdots,N$ are all independent and identical distributed as $t(3)$.
\end{itemize}

In this section, we consider two sample sizes, specifically $T=60$ and $T=120$, and three dimensions, namely $N=100$, $N=200$, and $N=400$. Our objective is to assess the empirical sizes of various tests under these four distinct scenarios. Table \ref{t4} presents the empirical sizes of each test for the aforementioned combinations of sample sizes and dimensions.

Upon examining the results, we observed that the least-squares-based tests--PY, MAX, and COM--experience size distortion as the dimension increases relative to the sample size. This phenomenon suggests that these tests may not effectively control the type I error rate when the dimensionality of the data is high compared to the sample size. Such size distortion can lead to inflated false positive rates, which is undesirable in statistical testing. In contrast, the spatial-sign-based procedures--SS, SM, and CC--demonstrate the ability to control the empirical sizes in most cases. This finding indicates that these tests are more robust to the dimensionality of the data and are better suited for handling high-dimensional datasets. By effectively controlling the type I error rate, these spatial-sign-based procedures offer a more reliable and accurate approach for testing in high-dimensional settings.

Moreover, the ability of the spatial-sign-based procedures to control the empirical sizes in high-dimensional settings is particularly crucial in fields such as finance and economics, where datasets with a large number of variables are common. In these fields, accurate and reliable statistical testing is essential for drawing valid conclusions and making informed decisions. Therefore, the use of spatial-sign-based procedures can greatly enhance the robustness and reliability of statistical analysis in high-dimensional settings. These findings are consistent with many previous works, such as \cite{wang2015high,Feng2016Multivariate,liu2023high,zhao2022high,zhao2023,zhao2023spatial}.

\begin{table}[!ht]
\begin{center}
\caption{\label{t4} Sizes of tests.}
                     \vspace{0.5cm}
                     \renewcommand{\arraystretch}{0.8}
                     \setlength{\tabcolsep}{7pt}{
\begin{tabular}{c|cc|cc|cc|cc}
\hline \hline
 Scenario & \multicolumn{2}{c}{{(I)}} & \multicolumn{2}{c}{{(II)}}& \multicolumn{2}{c}{{(III)}} & \multicolumn{2}{c}{{(IV)}}\\ \hline
$T$ &60&120&60&120&60&120&60&120\\
\hline
&\multicolumn{8}{c}{$N=100$}\\ \hline
PY&0.026&0.017&0.045&0.014&0.037&0.022&0.018&0.012\\
MAX&0.094&0.064&0.063&0.025&0.078&0.053&0.074&0.043\\
COM&0.066&0.04&0.06&0.015&0.054&0.033&0.047&0.031\\
SS&0.045&0.044&0.057&0.055&0.054&0.046&0.046&0.051\\
SM&0.04&0.055&0.052&0.04&0.061&0.055&0.04&0.051\\
CC&0.051&0.054&0.053&0.054&0.045&0.057&0.044&0.058\\ \hline
&\multicolumn{8}{c}{$N=200$}\\ \hline
PY&0.059&0.049&0.141&0.097&0.105&0.062&0.069&0.039\\
MAX&0.129&0.067&0.098&0.05&0.086&0.042&0.087&0.056\\
COM&0.097&0.06&0.143&0.076&0.118&0.045&0.083&0.044\\
SS&0.043&0.058&0.046&0.057&0.065&0.06&0.063&0.046\\
SM&0.046&0.05&0.053&0.059&0.056&0.047&0.06&0.051\\
CC&0.053&0.041&0.042&0.064&0.05&0.041&0.048&0.049 \\ \hline
&\multicolumn{8}{c}{$N=400$}\\ \hline
PY&0.137&0.134&0.228&0.203&0.212&0.145&0.158&0.096\\
MAX&0.151&0.087&0.088&0.041&0.1&0.066&0.09&0.043\\
COM&0.17&0.121&0.229&0.178&0.221&0.13&0.16&0.071\\
SS&0.042&0.053&0.064&0.059&0.056&0.06&0.055&0.041\\
SM&0.054&0.061&0.054&0.051&0.042&0.046&0.062&0.043\\
CC&0.058&0.056&0.041&0.06&0.064&0.06&0.04&0.065\\
\hline
\hline
\end{tabular}}
\end{center}
\end{table}

\begin{figure}[!ht]
\caption{Power of tests with different sparsity levels. \label{power}}
\centering
\subfloat[$(T,N)=(60,100)$]{\includegraphics[width=1\textwidth]{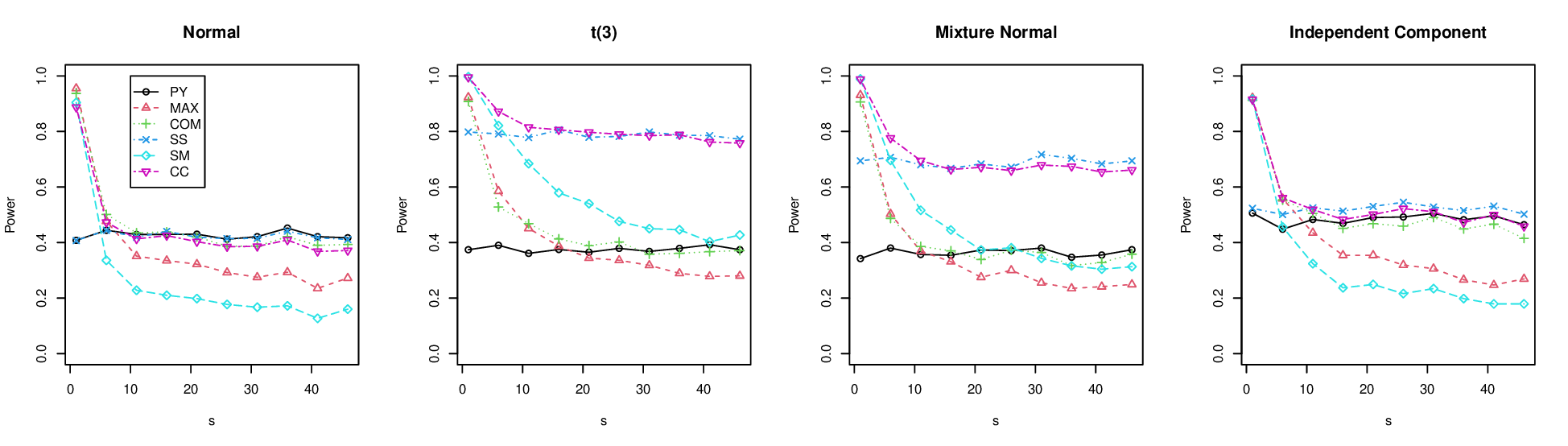}}\\
\subfloat[$(T,N)=(60,200)$]{\includegraphics[width=1\textwidth]{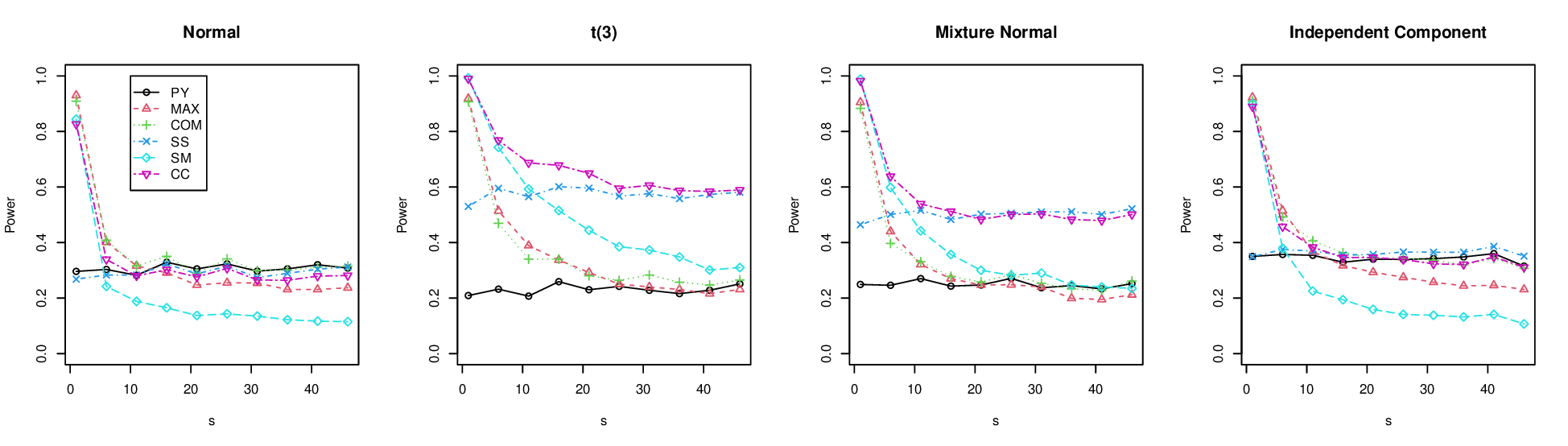}}
\end{figure}

\begin{figure}[!ht]
\caption{Power of tests with different signal levels over  $(T,N)=(60,200)$. \label{figsig}}
\subfloat[$s=2$]{\includegraphics[width=1\textwidth]{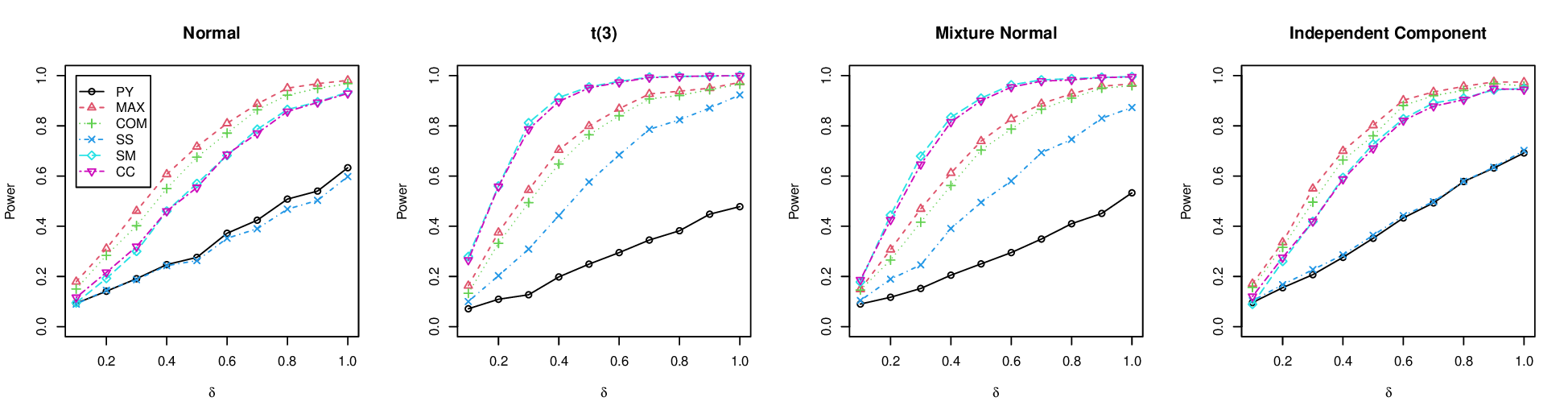}}\\
\subfloat[$s=10$]{\includegraphics[width=1\textwidth]{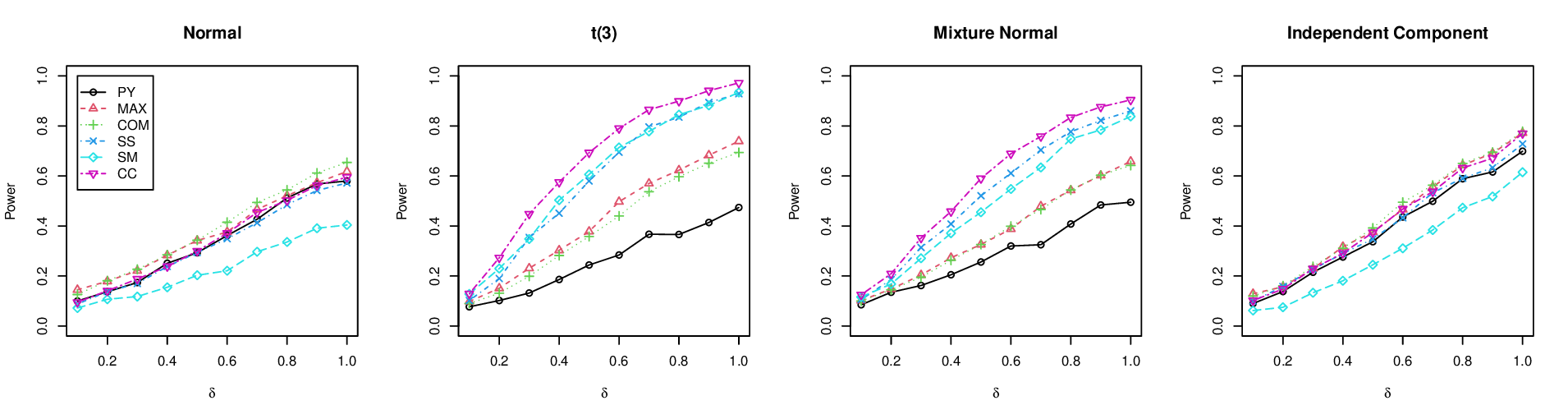}}\\
\subfloat[$s=25$]{\includegraphics[width=1\textwidth]{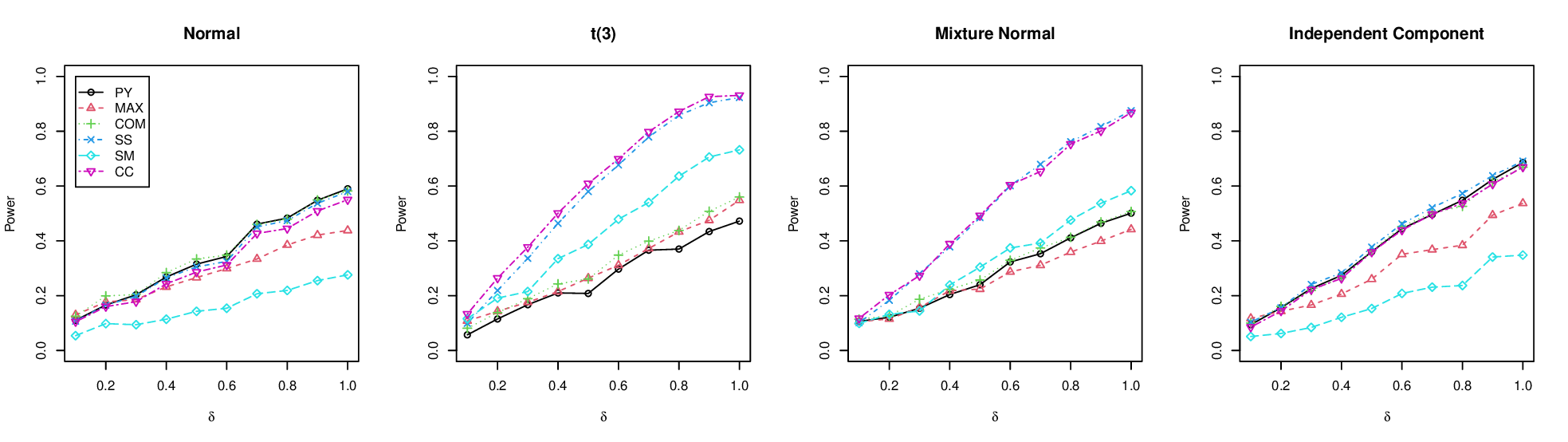}}
\end{figure}

For the alternative hypothesis, we set $\bm \alpha=(a,\cdots,a,0,\cdots,0)$, where the first $s$ components are equal to $a$. Figure \ref{power} displays the empirical power curves of each test with different sparsity levels, using $a=\sqrt{0.5/s}$ and considering $(T,N)=(60,100)$ and $(60,200)$. In the context of the multivariate normal distribution and the independent component model, the SS test exhibits similar performance to the PY test, which aligns with the findings in \cite{liu2023high}. Specifically, the SM test is slightly less powerful than the MAX test under these two models. However, when the errors exhibit heavy tails, such as under the multivariate t distribution or the multivariate mixture normal distribution, the spatial-sign-based test procedures—namely, SS and SM, ultimately prove more powerful than the PY and MAX tests. Additionally, the sum-type test procedures--PY and SS tests--demonstrate good performance when there are numerous nonzero alphas. Conversely, the max-type test procedures--MAX and SM tests--outperform the sum-type test procedures for sparse alternatives, i.e., when the number of nonzero alphas is very small. Lastly, our proposed CC test exhibits nearly the best performance in all cases, demonstrating robustness not only to heavy-tailed distributions but also to varying sparsity levels. Therefore, the CC test is a doubly robust test procedure for assessing $\alpha$ in high-dimensional linear factor pricing models. Furthermore, we also consider different signal strengths with $(T,N)=(60,200)$ for each test. Figure \ref{figsig} illustrates the power curves of each test under varying signal levels $a=\sqrt{\delta /s}$ and three sparsity levels ($s=2,10,25$). The power of all tests increases as the signal strengthens, indicating the consistency of each test. Moreover, we observe similar results to those presented in Figure \ref{power}. The CC test remains the most effective in most scenarios.

The CC test demonstrates notable advantages in controlling the type I error and maintaining statistical power. Firstly, in terms of controlling the type I error, the CC test, as a spatial-sign-based procedure, effectively controls the empirical size and maintains the desired type I error rate in most cases. This is crucial for statistical testing as it ensures accuracy and reliability, preventing erroneous conclusions due to type I errors. Secondly, in terms of statistical power, the CC test also exhibits its strengths. Since it does not rely on the assumptions of normal distribution and sparsity of the alternatives, the CC test is more robust when dealing with high-dimensional data. In summary, the CC test excels in both controlling the type I error and maintaining statistical power, making it a reliable and powerful statistical testing method suitable for high-dimensional alpha test in linear factor pricing model.

\section{Real data application}\label{Emp}

In this study, we first apply our methods to the Standard \& Poor's 500 index, utilizing the same datasets as those employed in \cite{feng2022high} and \cite{liu2023high}. To accommodate changes in the index composition over time, we compiled returns on all securities constituting the S\&P 500 index each month from January 2005 to November 2018. Given the evolving nature of the index, we focus on $N=374$ securities that were consistently included in the S\&P 500 index throughout this period. We describe these panel data using the Fama-French three-factor model, which is formulated as follows:
\begin{align}\label{3f}
Y_{it} = r_{it} - r_{ft} = \alpha_i + \beta_{i1} (r_{mt} - r_{ft}) + \beta_{i2} SMB_t + \beta_{i3} HML_t + \epsilon_{it},
\end{align}
where $i \in \{1, \ldots, N=374\}$ and $t \in \{\tau, \ldots, \tau+T-1\}$. Here, $r_{mt} - r_{ft}$ represents the market factor, $\tau$ is the starting date, and $T$ is the length of the test datasets. We obtained time series data on the safe rate of return and market factors from Ken French's data library. The one-month US treasury bill rate serves as the risk-free rate ($r_{ft}$), while the value-weighted return on all NYSE, AMEX, and NASDAQ stocks from CRSP is used as a proxy for the market return ($r_{mt}$). The calculations for $SMB_t$ and $HML_t$ are based on the average return of three small portfolios minus the average return of three big portfolios, and the average return of two value portfolios minus the average return of two growth portfolios, respectively, using stocks listed on the NYSE, AMEX, and NASDAQ. $r_{it}$ denotes the return rate of security $i$ at time $t$. All data are measured in percent per month. As demonstrated in \cite{feng2022high}, the model (\ref{3f}) is suitable for modeling this real dataset, with residuals resembling white noise and no additional latent factors for security returns. And there are no change points in the regression parameters.

\begin{figure}[htbp]
\centering
\caption{$p$-value sequence of each test from 2005 to 2018 for US data. \label{figdata}}
\subfloat[$T=48$]{\includegraphics[width=0.85\textwidth]{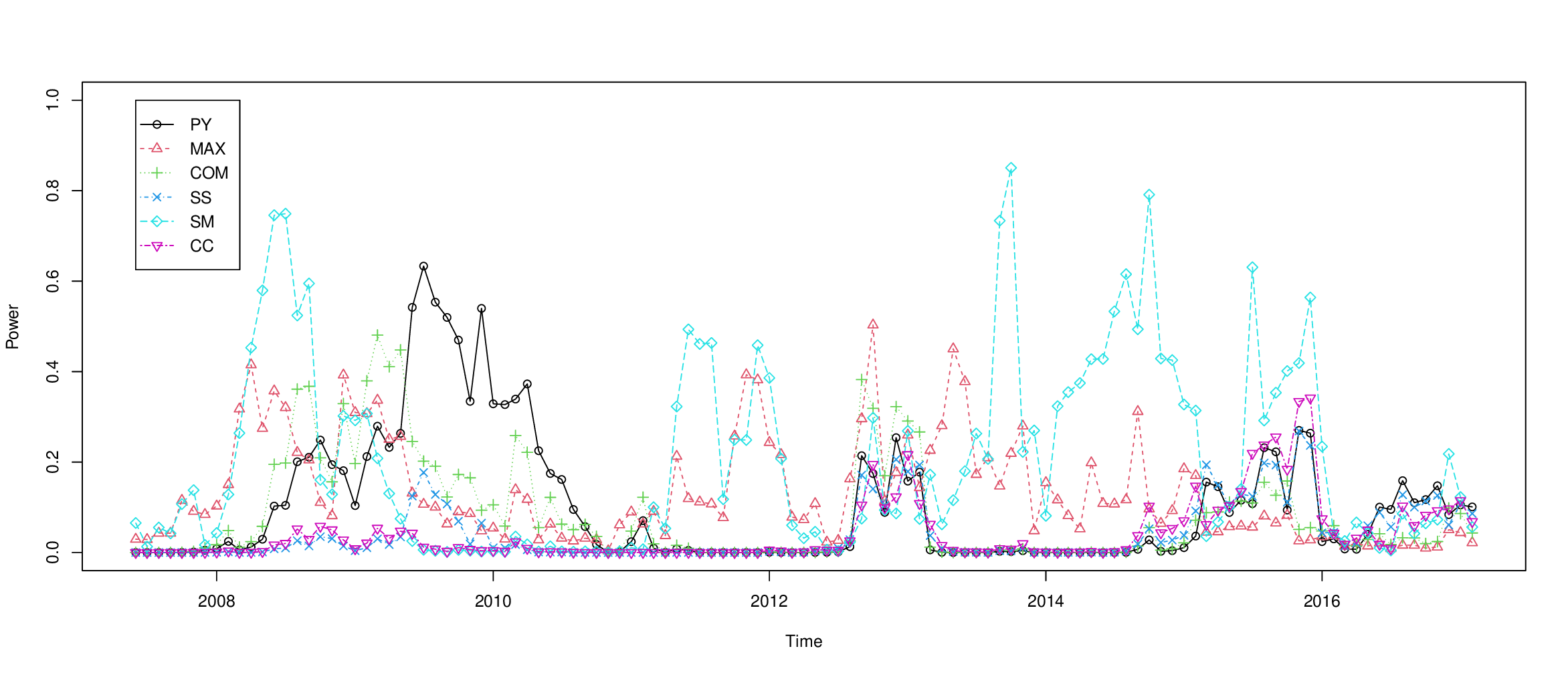}}\\
\subfloat[$T=60$]{\includegraphics[width=0.85\textwidth]{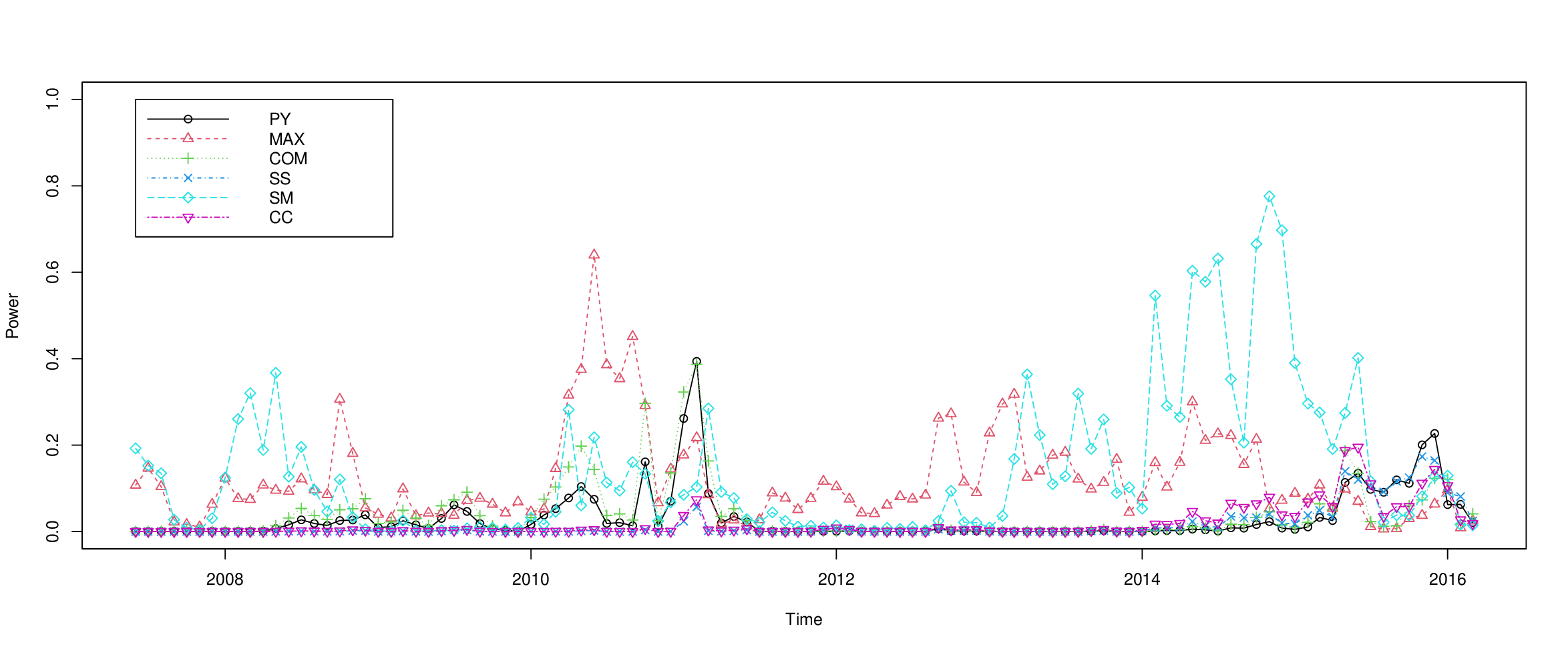}}\\
\subfloat[$T=72$]{\includegraphics[width=0.85\textwidth]{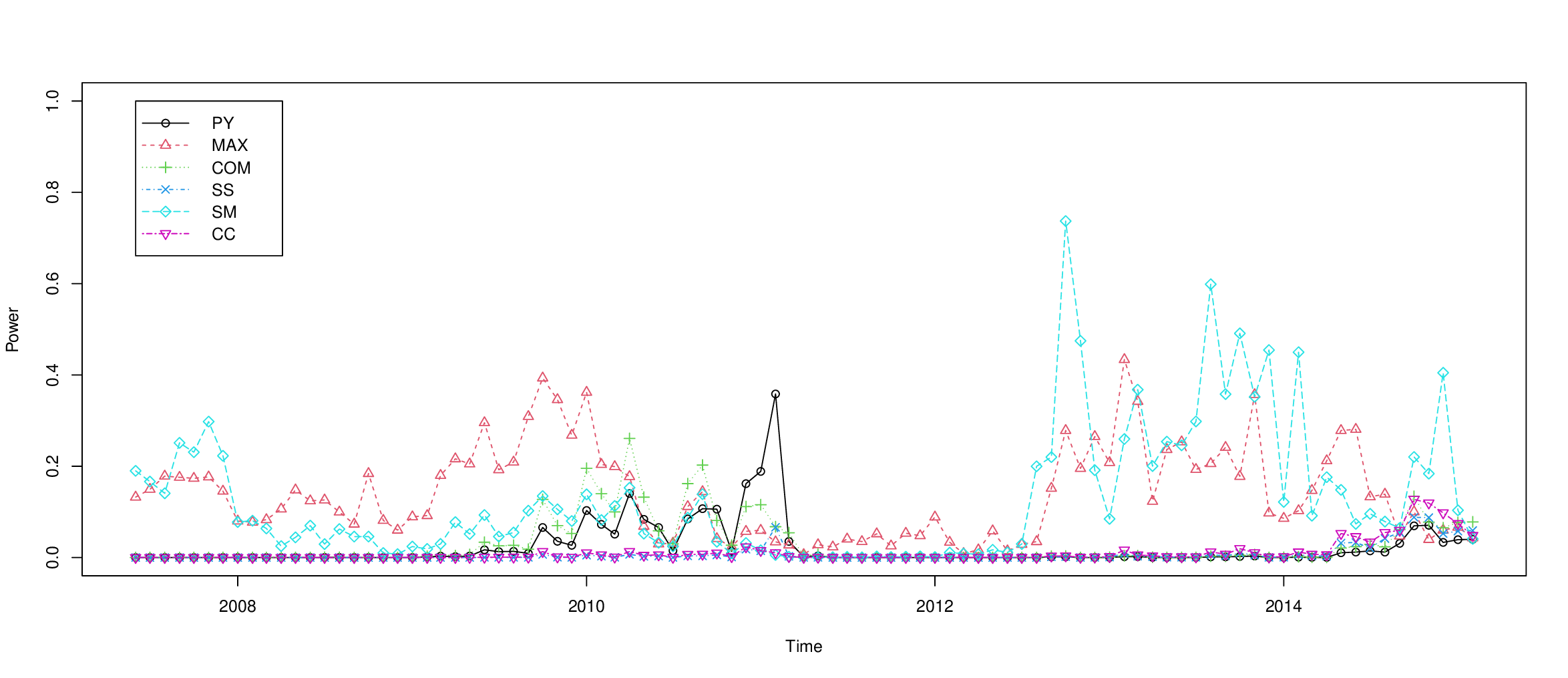}}
\end{figure}

Our focus is on testing the hypothesis:
\begin{align*}
H_0: \alpha_1 = \dots = \alpha_N = 0 \quad \text{versus} \quad H_1: \exists~i \in \{1, \ldots, N\} \text{ s.t. } \alpha_i \neq 0.
\end{align*}
For the entire dataset ($T=165$), the $p$-values of the PY, MAX, COM, SS, SM, and CC tests are 7.70e-09, 0.0654, 1.54e-08, 9.48e-09, 0.0298, 3.60e-08, respectively. These results provide strong evidence that $\bm{\alpha} \neq \bm{0}$. Therefore, we employ a rolling window approach, applying tests to observations within each of the $165-T$ rolling windows. Each rolling window corresponds to a length-$T$ consecutive subsequence of $\{1, \ldots, 165\}$, i.e., $\{\tau, \ldots, \tau+T-1\}$ for some $\tau \in \{1, \ldots, 166-T\}$. Table \ref{t2} presents the rejection ratios of the tests at significance levels $\gamma = 0.01, 0.05$ and for three choices of $T = 48, 60, 72$. These values of $T$ correspond to rolling windows covering 4, 5, or 6 years, respectively. Figure \ref{figdata} show the $p$-value sequence of each test from 2005 to 2018 for $T=48,60,72$, respectively.  Table \ref{t2} reveals that spatial-sign-based tests outperform corresponding least-squares-based tests, likely due to the non-normal distribution of residuals for most securities (see Figure S4 in \cite{liu2023high}). Additionally, sum-type test procedures exhibit better performance than max-type test procedures, which is unsurprising given the large number of nonzero alphas among the securities, as reported in \cite{feng2022high}. Finally, the newly proposed CC tests demonstrate nearly the best performance in most cases, highlighting the robustness of this method. These findings are consistent with theoretical results and simulation studies.


\begin{table}[!ht]
\begin{center}
\caption{\label{t2}  The rejection ratios of the six tests with three different window lengths.}
                     \vspace{0.5cm}
                     \renewcommand{\arraystretch}{0.8}
                     \setlength{\tabcolsep}{5pt}{
\begin{tabular}{c|cccccc|cccccc}
\hline \hline
 $T$ & \multicolumn{6}{c}{{$\gamma=0.01$}} & \multicolumn{6}{c}{{$\gamma=0.05$}}\\ \hline
 & PY&MAX&COM&SS&SM&CC & PY&MAX&COM&SS&SM&CC\\ \hline
 &\multicolumn{12}{c}{US data}\\ \hline
 48&0.436& 0.026& 0.359& 0.496& 0.162& 0.538&0.538& 0.282& 0.564& 0.709& 0.316& 0.726\\
60& 0.552& 0.029& 0.457& 0.771& 0.200& 0.733&0.810& 0.229& 0.752& 0.886& 0.448& 0.848\\
72&0.667& 0.022& 0.656& 0.860& 0.183& 0.774& 0.839& 0.237 &0.763& 0.925& 0.387& 0.925\\
\hline
\hline
\end{tabular}}
\end{center}
\end{table}

\section{Conclusion
}
In this paper, we address the problem of alpha testing in high-dimensional settings within the framework of linear factor pricing models. We first introduce a spatial-sign based max-type test procedure designed for sparse alternatives. By providing a Bahadur representation and Gaussian approximation for the spatial median estimator, as discussed in \cite{Feng2016Multivariate}, we establish the limiting null distribution and demonstrate the consistency of the proposed max-type test statistic. Subsequently, under some mild conditions, we show the asymptotic independence between this max-type test statistic and the existing sum-type test statistic which exhibits superior performance in the presence of dense alternatives. We then propose a robust Cauchy combination test, which performs exceptionally well in both sparse and dense scenarios as well as under heavy-tailed distributions. Our simulation studies and real data applications highlight the advantages and superior performance of the proposed Cauchy combination test procedure.
\appendix
\section{Proofs of all theorems}
First, we introduce Lemma 2 in \cite{liu2023high}.
\begin{lemma}\label{ddi}
	Under Conditions (C1)-(C4), we have $\max_{1\leq i\leq N}|\hat{d}_i^2-d_{i}^2|=O_p(\sqrt{\log N/T})$, where 
	$\hat{\D}=\diag\{\hat{d}_1^2,\dots,\hat{d}_N^2\}$ and ${\D}=\diag\{{d}_1^2,\dots,{d}_N^2\}$.
\end{lemma}
\begin{lemma}
	Under Conditions (C1)-(C5) and the null hypothesis, when $\min(T,N)\rightarrow \infty$, $T_{SS}\cd N(0,1)$.
\end{lemma}
\subsection{Proof of Theorem 1}
Under the null hypothesis, we have that $\bm \alpha=\mathbf{0}$ and $\Z_t=\bmv_{\cdot t}-\bmv^{\top}\X(\X^{\top}\X)^{-1}\f_t$. Define $\hat{\U}_t=U(\hat{\D}^{-1/2}\bmv_{\cdot t})$ and $\hat{r}_t=\|\hat{\D}^{-1/2}\bmv_{\cdot t}\|$
The estimator $\hat{\bm \theta}$ satisfies $\sum_{t=1}^TU(\hat{\D}^{-1/2}(\Z_t-\hat{\bm \theta}))=\mathbf{0}$, which is equivalent to 
\begin{align*}
	&\frac{1}{T}\sum_{t=1}^T(\hat{\U}_t-\hat{r}^{-1}_t\hat{\D}^{-1/2}\hat{\bm \theta}-\hat{r}^{-1}_t\hat{\D}^{-1/2}\bmv^{\top}\X(\X^{\top}\X)^{-1}\f_t)\\
	&\times(1+\hat{r}^{-2}_t\|\hat{\D}^{-1/2}\hat{\bm \theta}\|^2+\hat{r}^{-2}_t\|\hat{\D}^{-1/2}\bmv^{\top}\X(\X^{\top}\X)^{-1}\f_t\|^2\\
	&\quad\quad-2\hat{r}^{-1}_t\hat{\U}^\top_t\hat{\D}^{1/2}\hat{\bm \theta}-2\hat{r}^{-1}_t\hat{\U}^\top_t\hat{\D}^{1/2}\bmv^{\top}\X(\X^{\top}\X)^{-1}\f_t\\
	&\quad\quad+2\hat{r}^{-2}_t\hat{\bm \theta}^\top\hat{\D}^{-1}\bmv^{\top}\X(\X^{\top}\X)^{-1}\f_t)^{-1/2}=\mathbf{0}.
\end{align*}
Let ${\bm V}_t=\X(\X^{\top}\X)^{-1}\f_t$. Because $\f_t$ is independent of the errors $\varepsilon_{it}$, we have 
\begin{align}\label{hDetvt2}
	&E(\|\hat{\D}^{-1/2}\bmv^\top{\bm V}_t\|^2)\n\\
	=&E(\f_t^\top (\X^\top\X)^{-1}\X^\top\bmv\hat{\D}^{-1}\bmv^\top\X(\X^\top\X)^{-1}\f_t)\n\\
	=&O(N)E(\f_t^\top (\X^\top\X)^{-1}\X^\top 
	\I_{T}\X(\X^\top\X)^{-1}\f_t)\n\\
	=&O(NT^{-1}),
\end{align}
where the last second equality holds because $\bmv_{\cdot 1},\dots, \bmv_{\cdot T}$ are independent and identically distributed random vectors, and  $E(\bmv\hat{\D}^{-1}\bmv^\top)=E(\bmv{\D}^{-1}\bmv^\top)-E\{\bmv({\D}^{-1}-\hat{\D}^{-1})\bmv^\top\}=\{1+o_p(1)\}E(\varepsilon_{11}^2/d_1^2+\dots+\varepsilon_{1N}^2/d_N^2)\I_{T}=O(N)\I_T
$.
By the proof of Lemma A.3 in \cite{Feng2016Multivariate} and $\|\D^{-1/2}\bmv^{\top} \bm V_t\|^2=O_p(NT^{-1})$, we can obtain that $\|\hat{\bm \theta}\|=O_p(\zeta_1^{-1}T^{-1/2})$. According to the Taylor expansion, we can rewrite the above equation as 
\begin{align*}
	&\frac{1}{T}\sum_{t=1}^T(\hat{\U}_t-\hat{r}^{-1}_t\hat{\D}^{-1/2}\hat{\bm \theta}-\hat{r}^{-1}_t\hat{\D}^{-1/2}\bmv^{\top}\X(\X^{\top}\X)^{-1}\f_t)\\
	&\times(1-\frac{1}{2}\hat{r}^{-2}_t\|\hat{\D}^{-1/2}\hat{\bm \theta}\|^2-\frac{1}{2}\hat{r}^{-2}_t\|\hat{\D}^{-1/2}\bmv^{\top}\X(\X^{\top}\X)^{-1}\f_t\|^2\\
	&\quad\quad+\hat{r}^{-1}_t\hat{\U}^\top_t\hat{\D}^{1/2}\hat{\bm \theta}+\hat{r}^{-1}_t\hat{\U}^\top_t\hat{\D}^{1/2}\bmv^{\top}\X(\X^{\top}\X)^{-1}\f_t\\
	&\quad\quad-\hat{r}^{-2}_t\hat{\bm \theta}^\top\hat{\D}^{-1}\bmv^{\top}\X(\X^{\top}\X)^{-1}\f_t+\delta_{1t})=\mathbf{0},
\end{align*}
where $\delta_{1t}=O_p\big\{(-\frac{1}{2}\hat{r}^{-2}_t\|\hat{\D}^{-1/2}\hat{\bm \theta}\|^2-\frac{1}{2}\hat{r}^{-2}_t\|\hat{\D}^{-1/2}\bmv^{\top}{\bm V}_t\|^2+\hat{r}^{-1}_t\hat{\U}^\top_t\hat{\D}^{1/2}\hat{\bm \theta}+\hat{r}^{-1}_t\hat{\U}^\top_t\hat{\D}^{1/2}\bmv^{\top}{\bm V}_t-\hat{r}^{-2}_t\hat{\bm \theta}^\top\hat{\D}^{-1}\bmv^{\top}{\bm V}_t)^2\big\}$.
Moreover, 
$\hat{r}^{-2}_t=\|\hat{\D}^{-1/2}\bmv_{\cdot t}\|^{-2}
= \|{\D}^{-1/2}\bmv_{\cdot t}\|^{-2}(1+\| (\hat{\D}^{-1/2}-\D^{-1/2})\D^{1/2}\U_t\|^2-2\U_t^\top(\hat{\D}^{-1/2}-\D^{-1/2})\D^{1/2}\U_t)^{-1}={\color{black}\|{\D}^{-1/2}\bmv_{\cdot t}\|^{-2}}\times\{1+o_p(1)\}=O_p(N^{-1})$.
Hence, we have $\hat{r}^{-2}_t\|\hat{\D}^{-1/2}\bmv^{\top}\X(\X^{\top}\X)^{-1}\f_t\|^2=O(T^{-1})$.
Combined the above results and Condition (C3), we have $\delta_{1t}=O_p(T^{-1})$.
Then, we have
\begin{align}\label{eq1}
	&\frac{1}{T}\sum_{t=1}^T(1-\frac{1}{2}\hat{r}^{-2}_t\|\hat{\D}^{-1/2}\hat{\bm \theta}\|^2-\frac{1}{2}\hat{r}^{-2}_t\|\hat{\D}^{-1/2}\bmv^{\top}\X(\X^{\top}\X)^{-1}\f_t\|^2\\
	&-\hat{r}_t^{-2}\hat{\bm \theta}^\top\hat{\D}^{-1}\bmv^{\top}\X(\X^{\top}\X)^{-1}\f_t+\delta_{1t})\hat{\U}_t+\frac{1}{T}\sum_{t=1}^{T}(\hat{r}_t^{-1}\hat{\U}_t\hat{\U}_t^\top\hat{\D}^{-1/2}\hat{\bm\theta}+\hat{r}_t^{-1}\hat{\U}_t\hat{\U}_t^\top\hat{\D}^{-1/2}\bmv^\top{\bm V}_t)\n\\
	&-\frac{1}{T}\sum_{t=1}^{T}(1+\delta_{1t}+\delta_{2t})\hat{r}_t^{-1}\hat{\D}^{-1/2}\bmv^\top {\bm V}_t=	\frac{1}{T}\sum_{t=1}^{T}\hat{r}_t^{-1}(1+\delta_{1t}+\delta_{2t})\hat{\D}^{1/2}\hat{\bm\theta},
\end{align}
where \begin{align*}
	\delta_{2t}=&-\frac{1}{2}\hat{r}^{-2}_t\|\hat{\D}^{-1/2}\hat{\bm \theta}\|^2-\frac{1}{2}\hat{r}^{-2}_t\|\hat{\D}^{-1/2}\bmv^{\top}\X(\X^{\top}\X)^{-1}\f_t\|^2+\hat{r}^{-1}_t\hat{\U}^\top_t\hat{\D}^{1/2}\hat{\bm \theta}\n\\
	&+\hat{r}^{-1}_t\hat{\U}^\top_t\hat{\D}^{1/2}\bmv^{\top}\X(\X^{\top}\X)^{-1}\f_t-\hat{r}^{-2}_t\hat{\bm \theta}^\top\hat{\D}^{-1}\bmv^{\top}\X(\X^{\top}\X)^{-1}\f_t\\
	=&O_p(\delta_{1t}^{1/2}).
\end{align*}
Firstly, we focus on the first term on the left side of \eqref{eq1}. Due to Lemma \ref{ddi} and Equation (7.6) in \cite{liu2024spatial}, we can obtain that for any $j \in\{1,2, \cdots, p\}$,
\begin{align*}
	& \hat{U}_{t, j}-U_{t, j}=\frac{\left\|\D^{-1 / 2} \bmv_t\right\|}{\left\|\hat{\D}^{-1 / 2} \bmv_t\right\|} \cdot \frac{d_j}{\hat{d}_j} U_{t ,j}-U_{t ,j} \\
	& \leq(1+H)(1+H) U_{t ,j}-U_{t, j} \\
	&=H_u \U_{t ,j},
\end{align*}
where $\U_t=(U_{t,1},\dots,U_{t,N})$, $\hat{\U}_t=(\hat{U}_{t,1},\dots,\hat{U}_{t,N})$, $H=\max_{j}(d_{j}/\hat{d}_{j}-1)^2+\max_{j}(d_{j}/\hat{d}_{j}-1)=O_p(\sqrt{\log N/T})$, $H_u=O_p\left(H^2+2 H\right)=O_p\left(T^{-1 / 2}(\log p)^{1 / 2}\right) $.  Hence, $\hat{\U}_t-\U_t=H_u \U_t$. Similar to the Lemma 9 in \cite{liu2024spatial}, We can easily obtain
\begin{align*}
	&\left\|T^{-1} \sum_{t=1}^T \zeta_1^{-1} \hat{\U}_t\right\|_{\infty}\\
	\leq &|1+H_u|\left\|T^{-1} \sum_{t=1}^T \zeta_1^{-1} {\U}_t\right\|_{\infty}\\
	=&O_p\left\{T^{-1 / 2} \log ^{1 / 2}(NT)\right\}O_p\left\{1+\sqrt{\log N/T }\right\}\\
	=&O_p\left\{T^{-1 / 2} \log ^{1 / 2}(NT)\right\}.	
\end{align*}
Then, we calculate the second term. Similar the proof of Lemma A.4 in \cite{cheng2023}, we have  $\|\frac{1}{T}\sum_{t=1}^T\zeta_1^{-2}({\U}_t)^2\|_{\infty}=O_p(1)$, where $(\U_t)^2=(U_{t,1}^2,\dots,U_{t,N}^2)$. Then, we have
\begin{align*}
	& \left\|\zeta_1^{-1} T^{-1} \sum_{t=1}^T \hat{r}_t^{-2}\| \hat{\D}^{-1 / 2} \hat{\boldsymbol{\theta}}\|^2 \hat{\U}_t\right\|_{\infty} \leq\left|1+H_u\right| \cdot\left\|\zeta_1^{-1} T^{-1} \sum_{t=1}^T \hat{r}_t^{-2}\| \hat{\D}^{-1 / 2} \hat{\boldsymbol{\theta}}\|^2 \U_t\right\|_{\infty} \\
	& \leq \|\hat{\D}^{-1 / 2} \hat{\boldsymbol{\theta}}\|^2\sqrt{\left(T^{-1}\sum_{t=1}^T\hat{r}_t^{-4}\right)\|T^{-1}\sum_{t=1}^T\zeta_{1}^{-2}({\U}_t)^2\|_{\infty}} \left\{1+O_p\left(T^{-1 / 2}\log^{1 / 2} N\right)\right\}\\
	&=O_p\left(T^{-1}\right),
\end{align*}
where the last equality holds because $T^{-1}\sum_{t=1}^T\hat{r}_t^{-4}=\{1+o_p(1)\}T^{-1}\sum_{t=1}^T{r}_t^{-4}=O_p(\zeta_{1}^4)$ and $\|\hat{\D}^{-1 / 2} \hat{\boldsymbol{\theta}}\|^2=O_p(\zeta_{1}^{-2}T^{-1})$.
Then, let's focus on the third term. We have
\begin{align*}
	&\left\|\zeta_{1}^{-1}\frac{1}{T}\sum_{t=1}^{T}\hat{r}^{-2}_t\|\hat{\D}^{-1/2}\bmv^{\top}\X(\X^{\top}\X)^{-1}\f_t\|^2\hat{\U}_{t}\right\|_{\infty}\\
	\leq&\{1+o_p(1)\}\left\|\zeta_{1}^{-1}\frac{1}{T}\sum_{t=1}^{T}\hat{r}^{-2}_t\|\hat{\D}^{-1/2}\bmv^{\top}\X(\X^{\top}\X)^{-1}\f_t\|^2{\U}_{t}\right\|_{\infty}\\
	\leq &\sqrt{\left(T^{-1}\sum_{t=1}^T\hat{r}_t^{-4}\|\hat{\D}^{-1/2}\bmv^{\top}\X(\X^{\top}\X)^{-1}\f_t\|^4\right)\|T^{-1}\sum_{t=1}^T\zeta_{1}^{-2}({\U}_t)^2\|_{\infty}}
\{1+o_p(1)\}\\
	=&O_p(\zeta_{1}^{2})\max_{t}\|\hat{\D}^{-1/2}\bmv^{\top}\X(\X^{\top}\X)^{-1}\f_t\|^2\{1+o_p(1)\}.
\end{align*}
Next, we calculate $\max_{t}\|\hat{\D}^{-1/2}\bmv^{\top}\X(\X^{\top}\X)^{-1}\f_t\|^2$. Let $({\D}^{-1/2}\bmv^{\top})_{i}=(\varepsilon_{i1}/d_{i},\dots,\varepsilon_{iT}/d_{i})$ is the $i$-th row vector of ${\D}^{-1/2}\bmv^{\top}$.
Let $\bm{\Gamma}_{i}$ is the $i$-th row vector of $\mathbf{\Gamma}$.
Then, due to $\max_{1\leq i\leq N}\max_{1\leq s\leq T}\|W_{i,s}\|_{\psi_\alpha}\leq c_0$ for some constant $c_0>0$, we have $\|\bm{\Gamma}_{i}^\top \bm W_s\|_{\psi_\alpha}\lesssim c_0$ by Lemma B.4 in \cite{koike2021notes}. Then, according to the sub-exponential properties, we can obtain that $E(|\bm{\Gamma}_{i}^\top \bm W_s|^{k})\leq c_0^k(k)^{k/\alpha}\leq (c_0k)^{k}$ due to $1\leq \alpha\leq 2$ for any $k\leq 1$.
In addition, because of Conditions (C1)-(C4), we have $|V_{ts}|<K/T$ and  $E(|\varepsilon_{is}V_{ts}/d_i|^k)\leq (K\underline{d}^{-1}/T)^kE(v_s^k)E(|\bm{\Gamma}_{i}^\top\bm W_{t}|^k)\leq (K\underline{d}^{-1}/T)^kE(v_s^k) (c_0k)^k$, where $\V_t=(V_{t1},\dots,V_{tT})$.
By Stirling's approximation, we can obtain that $\varepsilon_{is}V_{ts}/d_i$
satisfies the condition of Bernstein. Then, according to the Bernstein inequality, we have 
\begin{align*}
	&P\left\{\max_{1\leq i\leq N}\max_{1\leq t\leq T}|({\D}^{-1/2}\bmv^\top)_i^{\top}\bm V_t|> CT^{-1/2}\log^{1/2}(NT)\right\}\\
	=&NTP\left\{|({\D}^{-1/2}\bmv^\top)_i^{\top}\bm V_t|>CT^{-1/2}\log^{1/2}(NT)\right\}\\
	=&NTP\left(|\sum_{s=1}^T\varepsilon_{is}V_{ts}/d_i|>CT^{-1/2}\log^{1/2}(NT)\right)\\
	=&2\exp\{-C^2\log(NT)\}NT\rightarrow 0,
\end{align*}
where $C$ is a sufficiently large constant. Hence, we have
\begin{align*}
	&\max_{t}\|\hat{\D}^{-1/2}\bmv^{\top}\X(\X^{\top}\X)^{-1}\f_t\|^2\\
	\leq &\max_{t}\|{\D}^{-1/2}\bmv^{\top}\X(\X^{\top}\X)^{-1}\f_t\|^2\{1+O_p(\sqrt{\log N/T})\}\\
	\leq &N\max_{1\leq i\leq N}\max_{1\leq t\leq T}|({\D}^{-1/2}\bmv^\top)_i^{\top}\bm V_t|^2\\
	\leq &NT^{-1}\log(NT).
\end{align*} 
and conclude that 
$$\left\|\zeta_{1}^{-1}\frac{1}{T}\sum_{t=1}^{T}\hat{r}^{-2}_t\|\hat{\D}^{-1/2}\bmv^{\top}\X(\X^{\top}\X)^{-1}\f_t\|^2\hat{\U}_{t}\right\|_{\infty}=O_p\{T^{-1}\log(NT)\}.$$
Similarly, we can obtain that 
$$\left\|\zeta_{1}^{-1}\frac{1}{T}\sum_{t=1}^{T}\hat{r}^{-1}_t\hat{\bm \theta}^{\top}\hat{\D}^{-1}\bmv^{\top}\X(\X^{\top}\X)^{-1}\f_t\hat{\U}_{t}\right\|_{\infty}=O_p\{T^{-1}\log(NT)\}.$$
Afterwards, we consider the fifth term. By Condition (C2) and Markov inequality, for any $\epsilon>0$, we have
\begin{align}\label{simarli}
	&P(\max_{1\leq t\leq T}r_{t}^{-1}\leq \epsilon\zeta_1T^{1/4})\n\\
	=&P(\max_{1\leq t\leq T}r_{t}^{-4}\leq \epsilon^4\zeta_1^4T)\n\\
	\leq&TE(r_{t}^{-4}) /(\epsilon^4\zeta_1^4T)\leq C\epsilon^{-4},	
\end{align}
where $C$ is some positive constant. Thus, $\max_{t}\hat{r}_t^{-2}=O_p(\zeta_1^{2}T^{1/2})$. Then, we can derive that  $\max_{t}\delta_{1t}=O_p(T^{-1/2})$ and $\max_{t}\delta_{2t}=O_p(T^{-1/4})$.
Hence, we have $\left\|\zeta_1^{-1} T^{-1} \sum_{t=1}^T \delta_{1 t} \hat{\U}_t\right\|_{\infty}=O_p\left(T^{-1}\right)$.
Next, we are concerned about the sixth term. Similar to proof of Lemma 8 in \cite{liu2024spatial}, we have
\begin{align*}
	& \left\|\zeta_1^{-1}\frac{1}{T}\sum_{t=1}^{T}\hat{r}_t^{-1}\hat{\U}_t\hat{\U}_t^\top\hat{\D}^{-1/2}\hat{\bm\theta}\right\|_{\infty} \\
	=&\left\|\zeta_1^{-1}\frac{1}{T}\sum_{t=1}^{T}{r}_t^{-1}{\U}_t{\U}_t^\top\hat{\D}^{-1/2}\hat{\bm\theta}\right\|_{\infty}\{1+o(1)\}\\
	\leqslant &\zeta_1^{-1} \|\frac{1}{T}\sum_{t=1}^{T}{r}_t^{-1}{\U}_tt{\U}_t^\top\|_1\left\|\hat{\D}^{-1 / 2} \hat{\boldsymbol{\theta}}\right\|_{\infty}\{1+o(1)\} \\
	\lesssim & N^{-1}\|\boldsymbol{\R}\|_1\left\|\hat{\D}^{-1 / 2} \hat{\boldsymbol{\theta}}\right\|_{\infty}\{1+o(1)\} \\
	+ & O_p\left( T^{-1 / 2} + N^{-1 / 6}+ N^{-\delta / 2}+\ T^{-1 / 2}(\log N)^{1 / 2}\left(N^{-3 / 2}+N^{-\delta / 2}\right)\right)\left\|\hat{\D}^{-1 / 2} \hat{\boldsymbol{\theta}}\right\|_{\infty}\{1+o(1)\} \\
 +&O_p\left( T^{-1 / 2} + N^{-1 / 6}+ N^{-\delta / 2}+ T^{-1 / 2}(\log N)^{1 / 2} N^{-3 / 2}\right)\left\|\hat{\D}^{-1 / 2} \hat{\boldsymbol{\theta}}\right\|_{\infty}\{1+o(1)\}.
\end{align*}
Then, we calculate the seventh term. We can obtain that 
{ \begin{align*}
	&\|\zeta_{1}^{-1}\frac{1}{T}\sum_{t=1}^T\hat{r}_t^{-1}\hat{\U}_t\hat{\U}_t^\top\hat{\D}^{-1/2}\bmv^\top\bm V_{t}\|_{\infty}\\
	=&\max_{1\leq i\leq N}\Big|\zeta_1^{-1}\frac{1}{T}\sum_{t=1}^T r_t^{-1}\sum_{j=1}^N\sum_{s=1}^Tr_sU_{s,j}U_{t,j}U_{t,i}V_{ts}\Big|\{1+o(1)\}\\
	\leq &\max_{1\leq i\leq N}\sqrt{\zeta_1^{-2}\frac{1}{T}\sum_{t=1}^T U_{t,i}^2}\sqrt{
		 \frac{1}{T}\sum_{t=1}^Tr_t^{-2}\Big(\sum_{j=1}^N\sum_{s=1}^Tr_sU_{s,j}U_{t,j}V_{ts}\Big)^2}\{1+o(1)\}\\
		 =&O_p(1)\sqrt{
		 	\frac{1}{T}\sum_{t=1}^Tr_t^{-2}\Big(\sum_{j=1}^N\sum_{s=1}^Tr_sU_{s,j}U_{t,j}V_{ts}\Big)^2}.
	\end{align*}
Because 
	\begin{align*}
	&E\frac{1}{T}\sum_{t=1}^Tr_t^{-2}\Big(\sum_{j=1}^N\sum_{s=1}^Tr_sU_{s,j}U_{t,j}V_{ts}\Big)^2\\
	=&E\frac{1}{T}\sum_{t=1}^Tr_t^{-2}\sum_{j_1=1}^N\sum_{s_1=1}^T\sum_{j_2=1}^N\sum_{s_2=1}^Tr_{s_1}U_{s_1,j_1}U_{t,j_1}V_{ts_1}r_{s_2}U_{s_2,j_2}U_{t,j_2}V_{ts_2}\\
	=&E\frac{1}{T}\sum_{t=1}^Tr_t^{-2}\sum_{j_1=1}^N\sum_{j_2=1}^Nr_{t}U_{t,j_1}U_{t,j_1}V_{tt}r_{t}U_{t,j_2}U_{t,j_2}V_{tt}\\
	+&E\frac{1}{T}\sum_{t=1}^Tr_t^{-2}\sum_{j_1=1}^N\sum_{s=1,s\neq t}^T\sum_{j_2=1}^Nr_{s}U_{s,j_1}U_{t,j_1}V_{ts}r_{s}U_{s,j_2}U_{t,j_2}V_{ts}\\
	=&O(T^{-2}+N^{-\delta}T^{-1}),
	\end{align*}
	where the last inequality holds because $\max _{j=1,\cdots,N}\sum_{\ell=1}^N\left|\sigma_{j \ell}\right| \leqslant a_0(N)=N^{1-\delta}$, $E(U_{t,i}U_{t,j})=N^{-1}\sigma_{i,j}+O(N^{-1-\delta/2})$ by Lemma A.4(iii) in \cite{cheng2023}.
	So, we can conclude that $\|\zeta_{1}^{-1}\frac{1}{T}\sum_{t=1}^T\hat{r}_t^{-1}\hat{\U}_t\hat{\U}_t^\top\hat{\D}^{-1/2}\bmv^\top\bm V_{t}\|_{\infty}=O_p(\max\{N^{-\delta/2}T^{-1/2},T^{-1}\})$.}
	
Finally, we consider $\left\|\zeta_{1}^{-1} \frac{1}{T}\sum_{t=1}^{T}(1+\delta_{1t}+\delta_{2t})\hat{r}_t^{-1}\hat{\D}^{-1/2}\bmv^\top\bm V_t\right\|_{\infty}$. We have
\begin{align*}
	&\zeta_{1}^{-1}\left\| \frac{1}{T}\sum_{t=1}^{T}\hat{r}_t^{-1}\hat{\D}^{-1/2}\bmv^\top\bm V_t\right\|_{\infty}\\
	=&\zeta_{1}^{-1}\left\| \frac{1}{T}\sum_{t=1}^{T}{r}_t^{-1}{\D}^{-1/2}\bmv^\top {\bm V}_t\right\|_{\infty}\{1+O_p(\sqrt{\log N /T})\}\\
	=&\zeta_{1}^{-1}\max_{1\leq i\leq N}\left|T^{-1}\sum_{t=1}^Tr_t^{-1}({\D}^{-1/2}\bmv^\top)_i^{\top}{\bm V}_t\right|\{1+O_p(\sqrt{\log N /T})\}\\
	\leq &\zeta_{1}^{-1}\max_{1\leq i\leq N}\max_{1\leq t\leq T}\left|({\D}^{-1/2}\bmv^\top)_i^{\top}{\bm V}_t\right|
	\left|T^{-1}\sum_{t=1}^Tr_t^{-1}\right|\{1+O_p(\sqrt{\log N /T})\}\\
	=&\max_{1\leq i\leq N}\max_{1\leq t\leq T}\left|({\D}^{-1/2}\bmv^\top)_i^{\top}{\bm V}_t\right|\\
	=&O_p\{T^{-1/2}\log^{1/2}(NT)\},
\end{align*}
and
{ \begin{align*}
	&\zeta_{1}^{-1}\left\| \frac{1}{T}\sum_{t=1}^{T}(\delta_{1t}+\delta_{2t})\hat{r}_t^{-1}\hat{\D}^{-1/2}\bmv^\top\bm V_t\right\|_{\infty}\\
	=&\zeta_{1}^{-1}\left\| \frac{1}{T}\sum_{t=1}^{T}{r}_t^{-1}(\delta_{1t}+\delta_{2t}){\D}^{-1/2}\bmv^\top {\bm V}_t\right\|_{\infty}\{1+O_p(\sqrt{\log N /T})\}\\
	=&\zeta_{1}^{-1}\max_{1\leq i\leq N}\left|T^{-1}\sum_{t=1}^Tr_t^{-1}(\delta_{1t}+\delta_{2t})({\D}^{-1/2}\bmv^\top)_i^{\top}{\bm V}_t\right|\{1+O_p(\sqrt{\log N /T})\}\\
	\leq &\zeta_{1}^{-1}\max_{1\leq t\leq T}|\delta_{1t}+\delta_{2t}|
	\max_{1\leq i\leq N}\max_{1\leq t\leq T}\left|({\D}^{-1/2}\bmv^\top)_i^{\top}{\bm V}_t\right|
	\left|T^{-1}\sum_{t=1}^Tr_t^{-1}\right|\{1+O_p(\sqrt{\log N /T})\}\\
	=&O_p(T^{-1/4})\max_{1\leq i\leq N}\max_{1\leq t\leq T}\left|({\D}^{-1/2}\bmv^\top)_i^{\top}{\bm V}_t\right|\\
	=&O_p\{T^{-3/4}\log^{1/2}(NT)\}.
\end{align*}}
 Additionally, we focus on the first term on the right side of \eqref{eq1}.
 Because $r_1,\dots,r_T$ are independently and identically distributed random variables, we can derive that 
\begin{align*}
	&T^{-1}\sum_{t=1}^T\hat{r}_t^{-1}(1+\delta_{1i}+\delta_{2i})\\
	=&T^{-1}\sum_{t=1}^T{r}_t^{-1}\{1+O_p(\sqrt{\log N /T})\}\{1+O_p(T^{-1/4})\}\\
	=&\zeta_1\{1+O_p(\sqrt{\log N /T})+O_p(T^{-1/4})\}.
\end{align*}
Then, combined with the above results, we rewritten the equation (\ref{eq1}) as
\begin{align*}
		\left\|\hat{\mathbf{D}}^{-1 / 2} \hat{\boldsymbol{\theta}}\right\|_{\infty} \leq & O_p\left\{T^{-1 / 2} \log ^{1 / 2}(NT)\right\} +\zeta_1^{-1}\left\|\frac{1}{T}\sum_{t=1}^{T}\hat{r}_t^{-1}\hat{\U}_t\hat{\U}_t^\top \hat{\mathbf{D}}^{-1 / 2} \hat{\bm\theta}\right\|_{\infty}\\
		\lesssim & N^{-1} a_0(N)\left\|\hat{\mathbf{D}}^{-1 / 2} \hat{\boldsymbol{\theta}}\right\|_{\infty}+O_p\left\{T^{-1 / 2} \log ^{1 / 2}(NT)\right\} \\
		& +O_p\left(T^{-1 / 2}+N^{-(1 / 6 \wedge \delta / 2)}+T^{-1 / 2}(\log N)^{1 / 2} N^{-3 / 2}\right)\left\|\hat{\mathbf{D}}^{-1 / 2} \hat{\boldsymbol{\theta}}\right\|_{\infty} .	
\end{align*}
Hence, due to $a_0(N)\asymp N^{1-\delta}$, we can derive that $$\left\|\hat{\mathbf{D}}^{-1 / 2} \hat{\boldsymbol{\theta}}\right\|_{\infty}=O_p\left\{T^{-1 / 2} \log ^{1 / 2}(NT)\right\}.$$

Furthermore, we have
\begin{align*}
	&\zeta_1^{-1}\left\|\frac{1}{T}\sum_{t=1}^{T}\hat{r}_t^{-1}\hat{\U}_t\hat{\U}_t^\top \hat{\mathbf{D}}^{-1 / 2} \hat{\bm\theta}\right\|_{\infty}\\
	=&O_p\Big\{T^{-1 }\log ^{1 / 2}(NT)+N^{-(1 / 6 \wedge \delta / 2)}T^{-1 / 2} \log ^{1 / 2}(NT)\\
	&+T^{-1 }(\log N)^{1 / 2} N^{-3 / 2}\log ^{1 / 2}(NT)\Big\}.
\end{align*}
{ Note that \begin{align*}
	&\frac{1}{T}\sum_{t=1}^{T}\U_t-\frac{1}{T}r_t^{-1}\D^{-1/2}\bmv^{\top}\F(\F^\top\F)^{-1}\f_t\\
	=&\frac{1}{T}\sum_{t=1}^{T}(1-\sum_{s=1}^Tr_s^{-1}r_tV_{st})\U_t.
\end{align*}}
Then, we can conclude that 
\begin{align}\label{theta=}
T^{1 / 2} \hat{\mathbf{D}}^{-1 / 2}\hat{\boldsymbol{\theta}}=T^{-1 / 2} \zeta_1^{-1} \sum_{t=1}^T(1-\sum_{s=1}^Tr_s^{-1}r_tV_{st}) \U_t+C_T,
\end{align}
where
\begin{align*}
	\left\|C_T\right\|_{\infty}= & O_p\left(T^{-1 / 2} \log ^{1 / 2}(T N)+N^{-(1 / 6 \wedge \delta / 2)} \log ^{1 / 2}(T N)+T^{-1 / 2}(\log N)^{1 / 2} N^{-3 / 2} \log ^{1 / 2}(T N)\right) \\
	& +O_p\left(T^{-1 / 4} \log ^{1 / 2}(T N)+T^{-1 / 2}(\log N)^{1 / 2} \log ^{1 / 2}(T N)\right) \\
	= & O_p\left(T^{-1 / 4} \log ^{1 / 2}(T N)+N^{-(1 / 6 \wedge \delta / 2)} \log ^{1 / 2}(T N)+T^{-1 / 2}(\log N)^{1 / 2} \log ^{1 / 2}(T N)\right) .
\end{align*}
Let $L_{N, T}=T^{-1 / 4} \log ^{1 / 2}(NT)+N^{-(1 / 6 \wedge \delta / 2)} \log ^{1 / 2}(NT)+T^{-1 / 2}(\log N)^{1 / 2} \log ^{1 / 2}(NT)$. Then for any sequence $\eta_T \rightarrow \infty$ and any $t \in \mathbb{R}^p$,
\begin{align*}
	P\left(T^{1 / 2} \hat{\mathbf{D}}^{-1 / 2}(\hat{\boldsymbol{\theta}}-\boldsymbol{\theta}) \leq t\right)  =&P\left(T^{-1 / 2} \zeta_1^{-1} \sum_{t=1}^T (1-\sum_{s=1}^Tr_s^{-1}r_tV_{st})\boldsymbol{U}_t+C_T \leq t\right) \\
	 \leq &P\left(T^{-1 / 2} \zeta_1^{-1} \sum_{t=1}^T (1-\sum_{s=1}^Tr_s^{-1}r_tV_{st})\boldsymbol{U}_t \leq t+\eta_T L_{N, T}\right)\\
	&+P\left(\left\|C_T\right\|_{\infty}>\eta_T L_{N, T}\right).
\end{align*}
{ When $r_{1},\dots,r_{N},\f_1\dots,\f_{T}$ are given, according to Lemma A4 in \cite{cheng2023} and the Gaussian approximation for independent partial sums in \cite{koike2021notes}, let $$\boldsymbol{G} \sim N\left(0, \zeta_1^{-2} T^{-1 } \sum_{t=1}^T (1-\sum_{s=1}^Tr_s^{-1}r_tV_{st})^2\mathbb{E}\left(\boldsymbol{U}_1 \boldsymbol{U}_1^{\top}\right)\right),$$ we can derive that the the conditional probability is
\begin{align*}
	&P_1\left(T^{1 / 2} \zeta_1^{-1} \sum_{t=1}^T (1-\sum_{s=1}^Tr_s^{-1}r_tV_{st}) \boldsymbol{U}_t \leq t+\eta_T L_{N, T}\right)\\
	 \leq  &P_1\left(\boldsymbol{G} \leq t+\eta_T L_{N, T}\right)+O\left(\left\{T^{-1} \log ^5(NT)\right\}^{1 / 6}\right) \\
	 \leq &P_1(\boldsymbol{G} \leq t)+O\left\{\eta_T L_{N, T} \log ^{1 / 2}(N)\right\}+O\left(\left\{T^{-1} \log ^5(NT)\right\}^{1 / 6}\right),
\end{align*}
where the second inequality holds due to Lemma 12 in \cite{liu2024spatial}. Note that \begin{align*}
	&T^{-1 } \sum_{t=1}^T (1-\sum_{s=1}^Tr_s^{-1}r_tV_{st})^2\\
    =&1-\frac{2}{T}\sum_{t=1}^T\sum_{s=1}^{T}r_s^{-1}r_tV_{st}+\frac{1}{T}\sum_{t=1}^T\sum_{s=1}^{T}r_s^{-2}r_t^2V_{st}^2\\
    +&\frac{1}{T}\sum_{t=1}^T\underset{s_1\neq s_2}{\sum^T\sum^T}r_{s_1}^{-1}r_{s_2}^{-1}r_t^2V_{s_1t}V_{s_2t}
	\cp	1-2\eta E(r_t^{-1})E(r_t)+\eta E(r_s^{-2})E(r_t^2),
\end{align*} 
 where $T^{-1}\1_T^{\top}\F(\F^{\top}\F)^{-1}\F^{\top}\1_T\cp \eta$.
Thus, let $1-2\eta E(r_t^{-1})E(r_t)+\eta E(r_s^{-2})E(r_t^2)=\eta_{\omega}$ and $\boldsymbol G^{'}\sim N\left(0, \zeta_1^{-2}\eta_{\omega} \mathbb{E}\left(\boldsymbol{U}_1 \boldsymbol{U}_1^{\top}\right)\right),$ we have
\begin{align*}
	P\left(T^{1 / 2} \hat{\mathbf{D}}^{-1 / 2}(\hat{\boldsymbol{\theta}}-\boldsymbol{\theta}) \leq t\right) \leq & P(\boldsymbol{G}^{'} \leq t)+O\left\{\eta_T L_{N, T} \log ^{1 / 2}(N)\right\}+O\left(\left\{T^{-1} \log ^5(NT)\right\}^{1 / 6}\right) \\
	& +P\left(\left|C_T\right|_{\infty}>\eta_T L_{N, T}\right)+o(1).
\end{align*}}
On the other hand, we have
\begin{align*}
	P\left(T^{1 / 2} \hat{\mathbf{D}}^{-1 / 2}(\hat{\boldsymbol{\theta}}-\boldsymbol{\theta}) \leq t\right) \geq & P(\boldsymbol{G}^{'} \leq t)-O\left\{\eta_T L_{N, T} \log ^{1 / 2}(N)\right\}-O\left(\left\{T^{-1} \log ^5(NT)\right\}^{1 / 6}\right) \\
	& -P\left(\left\|C_T\right\|_{\infty}>\eta_T L_{N, T}\right)-o(1),
\end{align*}
where $P\left(\left\|C_T\right\|_{\infty}>\eta_T L_{N, T}\right) \rightarrow 0$ as $T \rightarrow \infty$.
Then we have that, if $\log N=o\left(T^{1 / 5}\right)$ and $\log T=o\left(N^{1 / 3 \wedge \delta}\right)$,
$$
\sup _{t \in \mathbb{R}^p}\left|P\left(T^{1 / 2} \hat{\mathbf{D}}^{-1 / 2}(\hat{\boldsymbol{\theta}}-\boldsymbol{\theta}) \leq t\right)-P(\boldsymbol{G}^{'} \leq t)\right| \rightarrow 0.
$$
Further,
$$
\rho_T\left(\mathcal{A}^{r e}\right)=\sup _{A \in \mathcal{A}^{r e}}\left|P\left(T^{1 / 2} \hat{\mathbf{D}}^{-1 / 2}(\hat{\boldsymbol{\theta}}-\boldsymbol{\theta}) \in A\right)-P(\boldsymbol{G}^{'} \in A)\right| \rightarrow 0,
$$
by the Corollary 5.1 in \cite{chernozhukov2017central}.
Let $\boldsymbol{Z}=(Z_1,\dots,Z_N) \sim N\left(0, N^{-1}\zeta_1^{-2}\eta_{\omega} \R\right)$.
By Condition (C3) and Theorem 2 in \cite{feng2022asymptotic}, we have $\mathbb{E} Z_j^2=\zeta_1^{-2}N^{-1}\eta_{\omega} \leq \bar{B}$ and $\mathbb{E}\left[\max _{1 \leq j \leq N} Z_j\right] \asymp(\sqrt{\log N+\log \log N})$. Let $$\Delta_0=\max _{1 \leq j, k \leq N}\left|N\left(\mathbb{E} \U_1 \U_1^{\top}\right)_{j, k}-\R_{j, k}\right|,$$ by Lemma A.4 in \cite{cheng2023}, we have
$$
\Delta_0=\max _{1 \leq j, k \leq N}\left|N\left(\mathbb{E} \U_1 \U_1^{\top}\right)_{j, k}-\mathbf{R}_{j, k}\right|=O\left(N^{-\delta / 2}\right) .
$$
According to Theorem 2 in \cite{chernozhukov2015comparison}, we get
$$
\sup _{t \in \mathbb{R}}\left|P\left(\|\boldsymbol{Z}\|_{\infty} \leqslant t\right)-P\left(\|\boldsymbol{G}^{'}\|_{\infty} \leqslant t\right)\right| \leqslant C^{\prime} T^{-1 / 3}(1 \vee \log (NT))^{2 / 3} \rightarrow 0.
$$
According to the Theorem 2 in \cite{feng2022asymptotic}, we have
$$
P\left(N \zeta_1^2 \eta_{\omega}^{-1}\max _{1 \leq i \leq N} Z_i^2-2 \log N+\log \log N \leq x\right) \rightarrow G(x)=\exp \left\{-\frac{1}{\sqrt{\pi}} e^{-x / 2}\right\},
$$
a cdf of the Gumbel distribution, as $N \rightarrow \infty$. Thus, according to Theorem 2 in \cite{feng2022asymptotic}.
\begin{align*}
	& \left|P\left(T||\hat{\D}^{-1/2}\hat{\bm \theta}||^2_\infty\zeta-2 \log N+\log \log N \leq x\right)-G(x)\right| \\
	\leq & \left|P\left(\zeta  T||\hat{\D}^{-1/2}\hat{\bm \theta}||^2_\infty-2 \log N+\log \log N \leq x\right)-P\left(\zeta^2 \max _{1 \leq i \leq N} Z_i^2-2 \log N+\log \log N \leq x\right)\right| \\
	& +\left|P\left( \zeta^2 \max _{1 \leq i \leq N} Z_i^2-2 \log N+\log \log N \leq x\right)-G(x)\right|+o(1) \rightarrow 0,
\end{align*}
for any $x \in \mathbb{R}$.
Lastly, we can obtain that desired conclusion.\hfill$\Box$

\subsection{Proof of Proposition 2.1}
Under the null hypothesis, we have
\begin{align*}
	\left\|\hat{\mathbf{D}}^{-1 / 2}\left(\Z_t-\hat{\boldsymbol{\theta}}\right)\right\|=r_t(1 & +r_t^{-2}\left\|\left(\hat{\mathbf{D}}^{-1 / 2}-\mathbf{D}^{-1 / 2}\right)\bmv_{\cdot  t}\right\|^2 +r_t^{-2}\left\|\hat{\mathbf{D}}^{-1 / 2}( \hat{\bm \theta}+
	\bmv^{\top}\bm V_t)\right\|^2\\
	& \left.+2  \U_t^{\top}\left(\hat{\mathbf{D}}^{-1 / 2}-\mathbf{D}^{-1 / 2}\right) \mathbf{D}^{1 / 2} \U_t\right) -2 r_t^{-1} \U_t^{\top} \hat{\mathbf{D}}^{-1 / 2} ( \hat{\bm \theta}+
	\bmv^{\top}\bm V_t)\\
	& \left.-2 r_t^{-1} \U_t \mathbf{D}^{1 / 2}\left(\dot{\mathbf{D}}^{-1 / 2}-\mathbf{D}^{-1 / 2}\right) \hat{\mathbf{D}}^{-1 / 2} ( \hat{\bm \theta}+
	\bmv^{\top}\bm V_t)\right)^{1 / 2}.
\end{align*}
Similarly, by Lemma \ref{ddi}, we can show that $$r_t^{-2} \|\left(\hat{\mathbf{D}}^{-1 / 2}-\mathbf{D}^{-1 / 2}\right)\bmv_{\cdot  t}\|^2=\|\left(\hat{\mathbf{D}}^{-1 / 2}-\mathbf{D}^{-1 / 2}\right)\D^{1/2}\U_t\|^2=O_p\left((\log N / T)^{1 / 2}\right)=o_p(1),
$$ Due to \eqref{hDetvt2} and $\|\hat{\bm \theta}\|=O_p(\zeta_1^{-1}T^{-1/2})$, we have $r_t^{-2}\left\| \hat{\mathbf{D}}^{-1 / 2} ( \hat{\bm \theta}+
\bmv^{\top}\bm V_t)\right\|^2=O_p\left(T^{-1}\right)=o_p(1)$ and by the Cauchy inequality, the other parts are also $o_p(1)$. So,
$$
T^{-1} \sum_{t=1}^T\left\|\hat{\mathbf{D}}^{-1 / 2}\left(\Z_t-\hat{\boldsymbol{\theta}}\right)\right\|^{-1}=\left(T^{-1} \sum_{t=1}^T\left\|\mathbf{D}^{-1 / 2}\bmv_{\cdot t}\right\|^{-1}\right)\left\{1+o_p(1)\right\} 
$$
and
$$
T^{-1} \sum_{t=1}^T\left\|\hat{\mathbf{D}}^{-1 / 2}\left(\Z_t-\hat{\boldsymbol{\theta}}\right)\right\|^{2}=\left(T^{-1} \sum_{t=1}^T\left\|\mathbf{D}^{-1 / 2}\bmv_{\cdot t}\right\|^{2}\right)\left\{1+o_p(1)\right\} .
$$ 
Because $r_1,\dots,r_T$ are independently and identically distributed random variables, we have { $$\left(T^{-1} \sum_{t=1}^T\left\|\mathbf{D}^{-1 / 2}\bmv_{\cdot t}\right\|^{2}\right)= E(r_t^2)\{1+o_p(1)\} \text{ and  }\zeta_1^{-1}\left(T^{-1} \sum_{t=1}^T\left\|\mathbf{D}^{-1 / 2}\bmv_{\cdot t}\right\|^{-1}\right)\cp 1.$$
 Obviously, $\var\left(T^{-1} \zeta_1^{-1} \sum_{t=1}^T r_t^{-1}\right)=O\left(T^{-1}\right)$ and $\var\left(T^{-1}  \sum_{t=1}^Tr_t^{2}\right)=O\left\{T^{-1}N^2\right\}$. Similarly, we can prove that 
 $$\left(T^{-1} \sum_{t=1}^T\left\|\mathbf{D}^{-1 / 2}\bmv_{\cdot t}\right\|\right)=E(r_t)+O_p(T^{-1/2}N^{1/2}) \text{ and  }\zeta_2^{-1}\left(T^{-1} \sum_{t=1}^T\left\|\mathbf{D}^{-1 / 2}\bmv_{\cdot t}\right\|^{-2}\right)\cp1.$$
 Similarly, we can derive that $\var\big(T^{-1}  \sum_{t=1}^T \zeta_2^{-1}r_t^{-2}\big)=O\big(T^{-1} \big)$, $\var\big(T^{-1}  \sum_{t=1}^T r_t\big)=O\big(T^{-1}N\big)$ and $T^{-1}\omega_T\cp 1-\eta.$}
 Finally, the proof is completed due to Slutsky's theorem.
 
       \hfill$\Box$
\subsection{Proof of Theorem 2}
Without loss of generality, we assume that $\|\bm \alpha\|_{\infty}=O(\sqrt{\log N /T})$ and $\Z_t=\bmv_{\cdot t}+(1-\eta_t)\bm \alpha-\bmv^{\top}\bm V_t$, where $\eta_t=\bm 1_{T}^{\top} \F(\F^{\top}\F)^{-1}\f_t$ and $\bm V_t=\F(\F^{\top}\F)^{-1}\f_t$. Recall that $T^{-1}\omega_T\cp 1-\eta$, obviously we have $\eta=E(\eta_t)$. Define $\omega=1-\eta$, $\hat{\U}_t=U(\hat{\D}^{-1/2}\bmv_{\cdot t})$ and $\hat{r}_t=\|\hat{\D}^{-1/2}\bmv_{\cdot t}\|$.
The estimator $\tilde{\bm \theta}=\hat{\bm \theta}-\omega\bm{\alpha}$ satisfies $\sum_{t=1}^TU(\hat{\D}^{-1/2}(\Z_t-\omega\bm{\alpha}-\title{\bm \theta}))=\mathbf{0}$, which is equivalent to 
\begin{align*}
	&\frac{1}{T}\sum_{t=1}^T(\hat{\U}_t-\hat{r}^{-1}_t\hat{\D}^{-1/2}\tilde{\bm\theta}-\hat{r}^{-1}_t\hat{\D}^{-1/2}\bmv^{\top}\bm V_t-\hat{r}_t^{-1}\hat{\D}^{-1/2}(\eta_t-\eta)\bm\alpha)\\
	&\times(1+\hat{r}^{-2}_t\|\hat{\D}^{-1/2}\tilde{\bm\theta}\|^2+\hat{r}^{-2}_t\|\hat{\D}^{-1/2}\bmv^{\top}\bm V_t\|^2-2\hat{r}^{-1}_t\hat{\U}^\top_t\hat{\D}^{-1/2}\tilde{\bm\theta}\\
	&\quad\quad-2\hat{r}^{-1}_t\hat{\U}^\top_t\hat{\D}^{-1/2}\bmv^{\top}\bm V_t+\hat{r}_t^{-2}\|\hat{\D}^{-1/2}(\eta_t-\eta)\bm\alpha\|^2-2\hat{r}_t^{-1}\hat{\U}_t^{\top}\hat{\D}^{-1/2}(\eta_t-\eta)\bm\alpha\\
	&\quad\quad+2\hat{r}^{-2}_t\tilde{\bm\theta}^\top\hat{\D}^{-1}\bmv^{\top}\bm V_t+2\hat{r}_t^{-2}\bm V^\top_t\bmv\hat{\D}^{-1}(\eta_t-\eta)\bm\alpha+2\hat{r}_t^{-2}\tilde{\bm\theta}^{\top}\hat{\D}^{-1}(\eta_t-\eta)\bm\alpha)^{-1/2}=\mathbf{0}.
\end{align*}
By the proof of Lemma A.3 in \cite{Feng2016Multivariate}, we can similarly obtain that $\|\tilde{\bm \theta}\|=O_p(\zeta_1^{-1}T^{-1/2})$. According to the Taylor expansion, we can rewrite the above equation as 
\begin{align*}
	&\frac{1}{T}\sum_{t=1}^T(\hat{\U}_t-\hat{r}^{-1}_t\hat{\D}^{-1/2}\tilde{\bm\theta}-\hat{r}^{-1}_t\hat{\D}^{-1/2}\bmv^{\top}\bm V_t-\hat{r}_t^{-1}\hat{\D}^{-1/2}(\eta_t-\eta)\bm\alpha)\\
	&\times(1-\frac{1}{2}\hat{r}^{-2}_t\|\hat{\D}^{-1/2}\tilde{\bm\theta}\|^2-\frac{1}{2}\hat{r}^{-2}_t\|\hat{\D}^{-1/2}\bmv^{\top}\bm V_t\|^2+\hat{r}^{-1}_t\hat{\U}^\top_t\hat{\D}^{-1/2}\tilde{\bm\theta}\\
	&\quad\quad+\hat{r}^{-1}_t\hat{\U}^\top_t\hat{\D}^{-1/2}\bmv^{\top}\bm V_t-\frac{1}{2}\hat{r}_t^{-2}\|\hat{\D}^{-1/2}(\eta_t-\eta)\bm\alpha\|^2+\hat{r}_t^{-1}\hat{\U}_t^{\top}\hat{\D}^{-1/2}(\eta_t-\eta)\bm\alpha\\
	&\quad\quad-\hat{r}^{-2}_t\tilde{\bm\theta}^\top\hat{\D}^{-1}\bmv^{\top}\bm V_t-\hat{r}_t^{-2}\bm V^\top_t\bmv\hat{\D}^{-1}(\eta_t-\eta)\bm\alpha-\hat{r}_t^{-2}\tilde{\bm\theta}^{\top}\hat{\D}^{-1}(\eta_t-\eta)\bm\alpha+\delta_{3t})=\mathbf{0},
\end{align*}
where $\delta_{3t}=O_p\big\{(-\frac{1}{2}\hat{r}^{-2}_t\|\hat{\D}^{-1/2}\tilde{\bm\theta}\|^2-\frac{1}{2}\hat{r}^{-2}_t\|\hat{\D}^{-1/2}\bmv^{\top}{\bm V}_t\|^2+\hat{r}^{-1}_t\hat{\U}^\top_t\hat{\D}^{-1/2}\tilde{\bm\theta}+\hat{r}^{-1}_t\hat{\U}^\top_t\hat{\D}^{-1/2}\bmv^{\top}{\bm V}_t-\hat{r}^{-2}_t\tilde{\bm\theta}^\top\hat{\D}^{-1}\bmv^{\top}{\bm V}_t-\frac{1}{2}\hat{r}_t^{-2}\|\hat{\D}^{-1/2}(\eta_t-\eta)\bm\alpha\|^2+\hat{r}_t^{-1}\hat{\U}_t^{\top}\hat{\D}^{-1/2}(\eta_t-\eta)\bm\alpha-\hat{r}_t^{-2}\bm V^\top_t\bmv\hat{\D}^{-1}(\eta_t-\eta)\bm\alpha-\hat{r}_t^{-2}\tilde{\bm\theta}^{\top}\hat{\D}^{-1}(\eta_t-\eta)\bm\alpha)^2\big\}=O_p(T^{-1}\log N)$.
Then, we have
\begin{align}\label{eq4}
	&\frac{1}{T}\sum_{t=1}^T(1-\frac{1}{2}\hat{r}^{-2}_t\|\hat{\D}^{-1/2}\tilde{\bm\theta}\|^2-\frac{1}{2}\hat{r}^{-2}_t\|\hat{\D}^{-1/2}\bmv^{\top}\bm V_t\|^2-\frac{1}{2}\hat{r}_t^{-2}\|\hat{\D}^{-1/2}(\eta_t-\eta)\bm\alpha\|^2\n\\
	&-\hat{r}_t^{-2}\tilde{\bm\theta}^\top\hat{\D}^{-1}\bmv^{\top}\bm V_t-\hat{r}_t^{-2}\bm V^\top_t\bmv\hat{\D}^{-1}(\eta_t-\eta)\bm\alpha-\hat{r}_t^{-2}\tilde{\bm\theta}^{\top}\hat{\D}^{-1}(\eta_t-\eta)\bm\alpha+\delta_{3t})\hat{\U}_t\n\\
&+\frac{1}{T}\sum_{t=1}^{T}(\hat{r}_t^{-1}\hat{\U}_t\hat{\U}_t^\top\hat{\D}^{-1/2}\tilde{\bm\theta}+\hat{r}_t^{-1}\hat{\U}_t\hat{\U}_t^\top\hat{\D}^{-1/2}\bmv^\top{\bm V}_t+\hat{r}_t^{-1}\hat{\U}_t\hat{\U}_t^{\top}\hat{\D}^{-1/2}(\eta_t-\eta)\bm\alpha)\n\\
&-\frac{1}{T}\sum_{t=1}^{T}(1+\delta_{3t}+\delta_{4t})\hat{r}_t^{-1}\hat{\D}^{-1/2}\bmv^\top {\bm V}_t-\frac{1}{T}\sum_{t=1}^{T}(1+\delta_{3t}+\delta_{4t})\hat{r}_t^{-1}\hat{\D}^{-1/2}(\eta_t-\eta)\bm \alpha\n\\
	=&	\frac{1}{T}\sum_{t=1}^{T}\hat{r}_t^{-1}(1+\delta_{3t}+\delta_{4t})\hat{\D}^{-1/2}\tilde{\bm\theta},
\end{align}
where \begin{align*}
	\delta_{4t}=&-\frac{1}{2}\hat{r}^{-2}_t\|\hat{\D}^{-1/2}\tilde{\bm\theta}\|^2-\frac{1}{2}\hat{r}^{-2}_t\|\hat{\D}^{-1/2}\bmv^{\top}\bm V_t\|^2+\hat{r}^{-1}_t\hat{\U}^\top_t\hat{\D}^{-1/2}\tilde{\bm\theta}\n\\
	&+\hat{r}^{-1}_t\hat{\U}^\top_t\hat{\D}^{-1/2}\bmv^{\top}\bm V_t-\hat{r}^{-2}_t\tilde{\bm\theta}^\top\hat{\D}^{-1}\bmv^{\top}\bm V_t2\\
	&-\frac{1}{2}\hat{r}_t^{-2}\|\hat{\D}^{-1/2}(\eta_t-\eta)\bm\alpha\|^2+\hat{r}_t^{-1}\hat{\U}_t^{\top}\hat{\D}^{-1/2}(\eta_t-\eta)\bm\alpha\\
	&-\hat{r}_t^{-2}\bm V^\top_t\bmv\hat{\D}^{-1}(\eta_t-\eta)\bm\alpha-\hat{r}_t^{-2}\tilde{\bm\theta}^{\top}\hat{\D}^{-1}(\eta_t-\eta)\bm\alpha\\
	=&O_p(\delta_{3t}^{1/2}).
\end{align*}
Moreover, we have
\begin{align*}
	&\left\|\zeta_{1}^{-1}\frac{1}{T}\sum_{t=1}^{T}\hat{r}^{-2}_t\|\hat{\D}^{-1/2}(\eta_t-\eta)\bm\alpha\|^2\hat{\U}_{t}\right\|_{\infty}\\
	\leq&\left|1+H_u\right|\left\|\zeta_{1}^{-1}\frac{1}{T}\sum_{t=1}^{T}\hat{r}^{-2}_t\|\hat{\D}^{-1/2}(\eta_t-\eta)\bm\alpha\|^2{\U}_{t}\right\|_{\infty}\\
	\leq &\sqrt{\left(T^{-1}\sum_{t=1}^T\hat{r}_t^{-4}\|\hat{\D}^{-1/2}(\eta_t-\eta)\bm\alpha\|^4\right)\|T^{-1}\sum_{t=1}^T\zeta_{1}^{-2}({\U}_t)^2\|_{\infty}}
	\left|1+H_u\right|\\
	=&O_p(\zeta_{1}^{2})O_p(NT^{-1}\log N)\left|1+H_u\right|=O_p(T^{-1}\log N).
\end{align*}
Similarly, we can obtain that $\left\|\zeta_{1}^{-1}\frac{1}{T}\sum_{t=1}^{T}\hat{r}_t^{-2}\bm V^\top_t\bmv\hat{\D}^{-1}(\eta_t-\eta)\bm\alpha\hat{\U}_{t}\right\|_{\infty}=O_p(T^{-1}\log N)$ and $\left\|\zeta_{1}^{-1}\frac{1}{T}\sum_{t=1}^{T}
\hat{r}_t^{-2}\tilde{\bm\theta}^{\top}\hat{\D}^{-1}(\eta_t-\eta)\bm\alpha\hat{\U}_{t}\right\|_{\infty}=O_p(T^{-1}\log N)$. Next, it's easy to obtain
{ \begin{align*}
	&\|\zeta_{1}^{-1}\frac{1}{T}\sum_{t=1}^T\hat{r}_t^{-1}\hat{\U}_t\hat{\U}_t^{\top}\hat{\D}^{-1/2}(\eta_t-\eta)\bm\alpha\|_{\infty}\\
=&\|\zeta_{1}^{-1}\frac{1}{T}\sum_{t=1}^T{r}_t^{-1}{\U}_t{\U}_t^{\top}{\D}^{-1/2}(\eta_t-\eta)\bm\alpha\|_{\infty}\{1+O_p(T^{-1/2}\log N)\}\\
=&\max_{1\leq i\leq N}\Big|\zeta_{1}^{-1}\frac{1}{T}\sum_{t=1}^T{r}_t^{-1}(\eta_t-\eta){U}_{t,i}\sum_{j=1}^N{U}_{t,j}\alpha_j/d_j\Big|\{1+O_p(T^{-1/2}\log N)\}\\
\leq &\max_{1\leq i\leq N}\sqrt{\zeta_{1}^{-2}\frac{1}{T}\sum_{t=1}^T{U}_{t,i}^2}\sqrt{\frac{1}{T}\sum_{t=1}^T{r}_t^{-2}(\eta_t-\eta)^2\Big(\sum_{j=1}^N{U}_{t,j}\alpha_j/d_j\Big)^2}\\
=&O_p(1)\sqrt{\frac{1}{T}\sum_{t=1}^T{r}_t^{-2}(\eta_t-\eta)^2\Big(\sum_{j=1}^N{U}_{t,j}\alpha_j/d_j\Big)^2}
\end{align*}
and \begin{align*}
	&E\frac{1}{T}\sum_{t=1}^T{r}_t^{-2}(\eta_t-\eta)^2\Big(\sum_{j=1}^N{U}_{t,j}\alpha_j/d_j\Big)^2\\
	=&E\frac{1}{T}\sum_{t=1}^T{r}_t^{-2}(\eta_t-\eta)^2\sum_{j=1}^N{U}_{t,j}^2\alpha_j^2/d_j^2\\
	&+E\frac{1}{T}\sum_{t=1}^T{r}_t^{-2}(\eta_t-\eta)^2\underset{j_1\neq j_2}{\sum^N\sum^N}{U}_{t,j_1}\alpha_{j_1}/d_{j_1}{U}_{t,j_2}\alpha_{j_2}/d_{j_2}\\
	=&E\frac{1}{T}\sum_{t=1}^T\zeta_2E(\eta_t-\eta)^2\sum_{j=1}^N\{N^{-1}+O(N^{-1-\delta/2})\}\alpha_j^2/d_j^2\\
	&+E\frac{1}{T}\sum_{t=1}^T\zeta_2E(\eta_t-\eta)^2\underset{j_1\neq j_2}{\sum^N\sum^N}\{N^{-1}\sigma_{j_1j_2}+O(N^{-1-\delta/2})\}\alpha_{j_1}/d_{j_1}\alpha_{j_2}/d_{j_2}\\
	=&\frac{1}{T}\sum_{t=1}^T\zeta_2E(\eta_t-\eta)^2N^{-1}\bm\alpha^{\top}{\D}^{-1/2}\mathbf{R}{\D}^{-1/2}\bm\alpha+O_p(N^{-\delta/2}T^{-1}\log N)\\
= &O_p(N^{-1}T^{-1}\log N \lambda_{\max}(\mathbf{R}))+O_p(N^{-\delta/2}T^{-1}\log N)\\
=&o_p(T^{-1}/\log N),
\end{align*}
where the second inequality holds because of Lemma A.4 in \cite{cheng2023}. Hence, we can obtain that $\|\zeta_{1}^{-1}\frac{1}{T}\sum_{t=1}^T\hat{r}_t^{-1}\hat{\U}_t\hat{\U}_t^{\top}\hat{\D}^{-1/2}(\eta_t-\eta)\bm\alpha\|_{\infty}=o_p(\sqrt{T^{-1}/\log N })$. 
In addition,
\begin{align*}
	&\zeta_{1}^{-1}\left\|\frac{1}{T}\sum_{t=1}^{T}(1+\delta_{3t}+\delta_{4t})\hat{r}_t^{-1}\hat{\D}^{-1/2}(\eta_t-\eta)\bm \alpha\right\|_{\infty}\\
	=&\zeta_{1}^{-1}\left\| \frac{1}{T}\sum_{t=1}^{T}{r}_t^{-1}{\D}^{-1/2}(\eta_t-\eta)\bm \alpha\right\|_{\infty}\{1+O_p(\sqrt{\log N /T})\}\\
	=&\zeta_{1}^{-1}\underline{d}\|\bm\alpha\|_{\infty}\left|T^{-1}\sum_{t=1}^Tr_t^{-1}(\eta_t-\eta)\right|\{1+O_p(\sqrt{\log N /T})\}\\
	=&O_p\{T^{-1}\log(N)\}.
\end{align*} }
Hence, similar to \eqref{theta=}, we have 
\begin{align}\label{h1theata=}
T^{1 / 2} \hat{\mathbf{D}}^{-1 / 2}(\hat{\boldsymbol{\theta}}-\omega\bm\alpha)=T^{-1 / 2} \zeta_1^{-1} \sum_{t=1}^T (1-\sum_{s=1}^Tr_s^{-1}r_tV_{st})\U_t+C_T,
\end{align}
where { $\|C_T\|_{\infty}=o_p(1/\sqrt{\log N})$.}
Similar to the proof of Proposition 2.1, we can derive that $\hat{\zeta}\cp\zeta.$
Let $u_N(y)=y+2 \log N-\log \log N$, $T=T^{1/2}||\hat{\D}^{-1/2}\hat{\bm \theta}||_\infty\hat\zeta^{1/2}$, $T^c=T^{1/2}||\hat{\D}^{-1/2}(\hat{\bm \theta}-\omega\bm\alpha)||_\infty\hat\zeta^{1/2}$ and recall that $T_{SM}=T||\hat{\D}^{-1/2}\hat{\bm \theta}||^2_\infty\hat\zeta-2 \log N+\log \log N$.
According to \eqref{h1theata=}, we have
\begin{align*}
	P\left(T^c-2\log N+\log\log N\le x\right)\to \exp \left\{-\frac{1}{\sqrt{\pi}}e^{-x/2}\right\}.
	\end{align*}
It is clear that, $T \geq T^{1 / 2}\left\|\hat{\mathbf{D}}^{-1 / 2} \omega\bm\alpha\right\|_{\infty}\hat{\zeta}-T^c$. Define $q_{1-\alpha}$ to be the $1-\alpha$ quantile of $G(y)$.  Combined with Assumption 2 and Lemma 6, we get
\begin{align*}
	& P\left(T||\hat{\D}^{-1/2}\hat{\bm \theta}||^2_\infty\hat\zeta \geq q_{1-\alpha}+u_N(y) \right) \\
	\geq & P\left(T^{1 / 2}\left\|\hat{\mathbf{D}}^{-1 / 2} \omega\bm\alpha\right\|_{\infty} \cdot \hat{\zeta} -T^c \geq \sqrt{q_{1-\alpha}+u_N(y)} \right) \\
	= & P\left(T^c \leq T^{1 / 2}\left\|\hat{\mathbf{D}}^{-1 / 2} \omega\bm\alpha\right\|_{\infty} \cdot \hat{\zeta} -\sqrt{q_{1-\alpha}+u_N(y)} \right) \\
	\geq & P\left(T^c \leq T^{1 / 2}\left(\left\|\mathbf{D}^{-1 / 2} \omega\bm\alpha\right\|_{\infty}-\left\|\left(\hat{\mathbf{D}}^{-1 / 2}-\mathbf{D}^{-1 / 2}\right) \omega\bm\alpha\right\|_{\infty}\right) \cdot \hat{\zeta} -\sqrt{q_{1-\alpha}+u_N(y)} \right) \\
	\geq &  P\left(T^c \leq T^{1 / 2}\left\|\mathbf{D}^{-1 / 2} \omega\bm\alpha\right\|_{\infty} \cdot\left(1+O_p\left(T^{-1 / 2} \log ^{1 / 2}(N )\right)\right) \cdot \hat{\zeta} -\sqrt{q_{1-\alpha}+u_N(y)} \right) \rightarrow 1,
\end{align*}
if $\|\bm\alpha\|_{\infty} \geq \widetilde{C} T^{-1 / 2}\left\{\log N -2 \log \log (1-\alpha)^{-1}\right\}^{1 / 2}$ for some large enough constant $\widetilde{C}$. Similar to the proof of Lemma 6 in \cite{liu2024spatial}, we can derive that $\max_{1\leq i\leq N}|\hat{d}_i^2-d_{i}^2|=O_p(\sqrt{\log N/T})$. Hence, the last inequality holds since
\begin{align*}
	\left\|\left(\hat{\mathbf{D}}^{-1 / 2}-\mathbf{D}^{-1 / 2}\right) \bm\alpha\right\|_{\infty} & =\max _{i=1,2, \cdots, N } \frac{\hat{d}_i-d_i}{\hat{d}_i d_i} \omega\alpha_i \leq \max _{i=1,2, \cdots, N }\left|1-\frac{d_i}{\hat{d}_i}\right| \cdot\left\|\mathbf{D}^{-1 / 2} \omega\bm\alpha\right\|_{\infty} \\
	& \leq O_p\left(T^{-1 / 2} \log ^{1 / 2}(N )\right)\left\|\mathbf{D}^{-1 / 2} \omega\bm\alpha\right\|_{\infty}.
\end{align*}

\hfill$\Box$
\subsection{Proof of Theorem 3}
 To prove $T_{S S}$ and $T_{SM}$ are asymptotically independent, it suffices to show that: Under $H_0$,
\begin{align}\label{nullinde}
	P\left(\frac{T_{SS}}{\sqrt{2\tr(\R^2)}} \leq x, T||\hat{\D}^{-1/2}\hat{\bm \theta}||^2_\infty\hat\zeta-2 \log N+\log \log N \leq y\right) \rightarrow \Phi(x) \cdot \exp \left\{-\frac{1}{\sqrt{\pi}} e^{-y / 2}\right\} .
\end{align}
Recall that $u_N(y)=y+2 \log N-\log \log N$, and we rewrite equation \eqref{nullinde} as
$$
P\left(\frac{T_{SS}}{\sqrt{2\tr(\R^2)}} \leq x, T||\hat{\D}^{-1/2}\hat{\bm \theta}||^2_\infty\hat\zeta \leq u_N(y)\right) \rightarrow \Phi(x) \cdot \exp \left\{-\frac{1}{\sqrt{\pi}} e^{-y / 2}\right\} .
$$
By the proof of Theorem 1 in \cite{liu2023high}, we acquire
$$
T_{SS}=\sqrt{\frac{1}{2 \operatorname{tr}\left(\mathbf{R}^2\right)}} N \omega_T^{-1} \underset{t_1\neq t_2}{\sum\sum} h_{t_1} h_{t_2} \U_{t_1}^{\top}  \U_{t_2}+o_p\left(1\right),
$$
where $\omega_T=\bm 1_T^\top \P_{\X}\bm
1_T$. Combined with \eqref{theta=}, it suffice to show,
\begin{align}\label{eq2}
	& P\left(\sqrt{\frac{1}{2 \operatorname{tr}\left(\mathbf{R}^2\right)}} N \omega_T^{-1} \underset{t_1\neq t_2}{\sum\sum} h_{t_1} h_{t_2} \U_{t_1}^{\top} \U_{t_2}+o_p\left(1\right) \leq x,\right.\n\\
	& \left.\quad N\eta_{\omega}^{-1}\left\|T^{-1 / 2} \sum_{t=1}^T(1-\sum_{s=1}^Tr_s^{-1}r_tV_{st}) \U_t\right\|_{\infty}^2+O_p\left(L_{N, T}\right) \leq u_N(y)\right) \n\\
	& \rightarrow \Phi(x) \cdot \exp \left\{-\frac{1}{\sqrt{\pi}} e^{-y / 2}\right\},
\end{align}
where $L_{N, T}=T^{-1 / 4} \log ^{1 / 2}(T N)+N^{-(1 / 6 \wedge \delta / 2)} \log ^{1 / 2}(T N)+T^{-1 / 2}(\log N)^{1 / 2} \log ^{1 / 2}(T N)$. Then, we just need to prove \eqref{eq2}.

{ First, we will prove that the equation \eqref{eq2} holds if $\U_t$ follows the normal distribution. Similar to the proof of Theorem 3 in \cite{feng2022asymptotic},  let $\U_t=\left(U_{t,1}, \cdots, U_{t,N}\right)^T \sim N(\mathbf{0}, \boldsymbol{\Sigma}_{u})$. For any set $\Lambda=\left\{i_1, \cdots, i_d\right\}$ with $1 \leqslant i_1<\cdots<i_d \leqslant N$, write $\boldsymbol{U}_{t,\Lambda}=\left(\U_{t,i_1}, \cdots,\U_{t,i_d}\right)^T$ and $\boldsymbol{U}_{t,\Lambda^C}=\left(\U_{t,j_1}, \cdots, \U_{t,j_{p-d}}\right)^T$ where $j_1<\cdots<j_{p-d}$ and $\left\{j_1, \cdots, j_{p-d}\right\}=\{1,2,\dots,N\}\setminus \Lambda$. Define $\mathbf{H}_{t}=\boldsymbol{U}_{t,\Lambda^{C}}-\boldsymbol{\Sigma}_{u,21} \boldsymbol{\Sigma}_{u,11}^{-1} \boldsymbol{U}_{t,\Lambda}$ and $\mathbf{V}_{t}=\boldsymbol{\Sigma}_{u,21} \boldsymbol{\Sigma}_{u,11}^{-1} \boldsymbol{U}_{t,\Lambda}$. Note that
\begin{align*}
&N \omega_T^{-1} \underset{t_1\neq t_2}{\sum\sum} h_{t_1} h_{t_2} \U_{t_1}^{\top} \U_{t_2}\\
=&\|N^{1/2} \omega_T^{-1/2} \sum^T_{t=1} h_{t} \U_{t}\|^2-N\\
=&\|N^{1/2} \omega_T^{-1/2} \sum^T_{t=1} h_{t} \U_{t1,\Lambda}\|^2+\|N^{1/2} \omega_T^{-1/2} \sum^T_{t=1} h_{t} \mathbf{H}_{t}\|^2\\
&+
\|N^{1/2} \omega_T^{-1/2} \sum^T_{t=1} h_{t} \mathbf{V}_{t}\|^2+2N^{1/2} \omega_T^{-1/2} \sum^T_{t=1} h_{t} \mathbf{V}_{t}^{\top}N^{1/2} \omega_T^{-1/2} \sum^T_{s=1} h_{s} \mathbf{H}_{s}-N
\end{align*}
and $\omega_T^{-1/2} \sum^T_{t=1} h_{t} \U_{t}\sim N(\mathbf{0}, \boldsymbol{\Sigma}_{u})$. Hence, according to the proof of Lemma 5.8 in \cite{feng2022asymptotic}, under Condition (C5) and $\tr(N^2\mathbf{\Sigma}_u^2)=\tr(\mathbf{R}^2)\{1+o(1)\}$, for any $\epsilon>0$ there exists $t=t_N \rightarrow \infty$ such that
$$
\max _{\Lambda} P\left(\left|\Theta_{N, \Lambda}\right| \geqslant \epsilon \sqrt{2\tr(\mathbf{R}^2)}\right) \leqslant \frac{1}{N^t},
$$
as $N$ is sufficiently large, where $\Theta_{N, \Lambda}=\|N^{1/2} \omega_T^{-1/2} \sum^T_{t=1} h_{t} \U_{t1,\Lambda}\|^2+\|N^{1/2} \omega_T^{-1/2} \sum^T_{t=1} h_{t} \mathbf{V}_{t}\|^2+2N^{1/2} \omega_T^{-1/2} \sum^T_{t=1} h_{t} \mathbf{V}_{t}^{\top}N^{1/2} \omega_T^{-1/2} \sum^T_{s=1} h_{s} \mathbf{H}_{s}$, the maximum $\Lambda=\left\{i_1, \cdots, i_d\right\}$ runs over all oossible indices $i_1, \cdots, i_d$ with $1 \leqslant i_1<\cdots<i_d \leqslant N$. Hence, similar to the proof of Theorem 3 in \cite{feng2022asymptotic}, we can obtain the  desired conclusion.}  We then investigate the non-normal case. Let $\boldsymbol{\xi}_t=$ $\U_t \in \mR^N, i=1,2, \cdots, T$. For $\boldsymbol{z}=\left(z_1, \cdots, z_q\right)^{\top} \in \mR^q$, we consider a smooth approximation of the maximum function, namely,
$$
F_\beta(\boldsymbol{z}):=\beta^{-1} \log \left(\sum_{j=1}^q \exp \left(\beta z_j\right)\right)
$$
where $\beta>0$ is the smoothing parameter that controls the level of approximation. An elementary calculation shows that for all $z \in \mR^q$,
$$
0 \leq F_\beta(\boldsymbol{z})-\max _{1 \leq j \leq q} z_j \leq \beta^{-1} \log q .
$$
Define $\sigma_{SS}^2=2\tr\left(\R^2\right)$,
\begin{align*}
	W\left(\boldsymbol{x}_1, \cdots, \boldsymbol{x}_T\right) & =\frac{\left\|N^{1/2} \sum_{t=1}^T\omega_{T}^{-1/2} h_{t}\boldsymbol{x}_t\right\|_2^2-N}{\sqrt{2 \operatorname{tr}\left(\mathbf{R}^2\right)}} \\
	& =\frac{N\omega_{T}^{-1} \sum_{i \neq j}h_ih_j \boldsymbol{x}_i^{\top} \boldsymbol{x}_j}{\sqrt{2  \operatorname{tr}\left(\mathbf{R}^2\right)}}:=\frac{N\omega_{T}^{-1} \sum_{i \neq j}h_ih_j \boldsymbol{x}_i^{\top} \boldsymbol{x}_j}{\sigma_{SS}}, \\
	V\left(\boldsymbol{x}_1, \cdots, \boldsymbol{x}_T\right) & =\beta^{-1} \log \left(\sum_{j=1}^N \exp \left(\beta \sqrt{\frac{N}{T}} \sum_{i=1}^T(1-\sum_{s=1}^Tr_s^{-1}r_iV_{si}) \boldsymbol{x}_{i, j}\right)\right) .
\end{align*}
By setting { $\beta $ to satisfy $\log N=o(\beta) $}, equation \eqref{eq2} is equivalent to
$$
P\left(W\left(\boldsymbol{\xi}_1, \cdots, \boldsymbol{\xi}_N\right) \leq x, V\left(\boldsymbol{\xi}_1, \cdots, \boldsymbol{\xi}_N\right) \leq u_N(y)\right) \rightarrow \Phi(x) \cdot \exp \left\{-\frac{1}{\sqrt{\pi}} e^{-y / 2}\right\}.
$$
Suppose $\left\{\boldsymbol{Y}_1, \boldsymbol{Y}_2, \cdots, \boldsymbol{Y}_T\right\}$ are sample from $N\left(0, E \U_1^{\top} \U_1\right)$, and independent with $\U_1, \cdots, \U_T$ or write as $\left(\boldsymbol{\xi}_1, \cdots, \boldsymbol{\xi}_T\right)$. The key idea is to show that: $\left(W\left(\boldsymbol{\xi}_1, \cdots, \boldsymbol{\xi}_T\right), V\left(\boldsymbol{\xi}_1, \cdots, \boldsymbol{\xi}_T\right)\right)$ has the same limiting distribution as $\left(W\left(\boldsymbol{Y}_1, \cdots, \boldsymbol{Y}_T\right), V\left(\boldsymbol{Y}_1, \cdots, \boldsymbol{Y}_T\right)\right)$.

Let $l_b^2(\mathbb{R})$ denote the class of bounded functions with bounded and continuous derivatives up to order 3. It is known that a sequence of randon variables $\left\{Z_n\right\}_{n=1}^{\infty}$ converges weakly to a random variable $Z$ if and only if for every $f \in l_b^3(\mathbb{R}), E\left(f\left(Z_n\right)\right) \rightarrow E(f(Z))$.
It suffices to show that:
$$
E\left\{f\left(W\left(\boldsymbol{\xi}_1, \cdots, \boldsymbol{\xi}_T\right), V\left(\boldsymbol{\xi}_1, \cdots, \boldsymbol{\xi}_T\right)\right)\right\}-E\left\{f\left(W\left(\boldsymbol{Y}_1, \cdots, \boldsymbol{Y}_T\right), V\left(\boldsymbol{Y}_1, \cdots, \boldsymbol{Y}_T\right)\right)\right\} \rightarrow 0,
$$
for every $f \in l_b^3\left(\mathbb{R}^2\right)$ as $(N, T) \rightarrow \infty$. We just need to prove
\begin{align*}
&E\left\{f\left(W\left(\boldsymbol{\xi}_1, \cdots, \boldsymbol{\xi}_T\right), V\left(\boldsymbol{\xi}_1, \cdots, \boldsymbol{\xi}_T\right)\right)\right\}\\
&-E\left\{f\left(W\left(\boldsymbol{Y}_1, \cdots, \boldsymbol{Y}_T\right), V\left(\boldsymbol{Y}_1, \cdots, \boldsymbol{Y}_T\right)\right)\right\} \rightarrow 0.
\end{align*}
Note that $\{f_1,\dots, f_T,r_1,\dots,r_T\}$ are independent with $\boldsymbol{\xi}_1,\dots,\boldsymbol{\xi}_T,\boldsymbol{Y}_1,\dots,\boldsymbol{Y}_T$.
Therefore, let $ \mathcal{F}_{fr}=\sigma\left\{f_1,\dots, f_T,r_1,\dots,r_T\right\}$. We introduce $\widetilde{W}_d=W\left(\boldsymbol{\xi}_1, \cdots, \boldsymbol{\xi}_{d-1}, \boldsymbol{Y}_d, \cdots, \boldsymbol{Y}_T\right)$ and $\widetilde{V}_d=V\left(\boldsymbol{\xi}_1, \cdots, \boldsymbol{\xi}_{d-1}, \boldsymbol{Y}_d, \cdots, \boldsymbol{Y}_T\right)$ for $d=1, \cdots, T+1$, $\mathcal{F}_d=\sigma\left\{\boldsymbol{\xi}_1, \cdots, \boldsymbol{\xi}_{d-1}, \boldsymbol{Y}_{d+1}, \cdots, \boldsymbol{Y}_T\right\}$ for $d=1, \cdots, T$. If there is no danger of confusion, we simply write $\widetilde{W}_d$ and $\widetilde{V}_d$ as $W_d$ and $V_d$ respectively (only for this part). Then,
{ \begin{align*}
	& \left|E\left\{f\left(W\left(\boldsymbol{\xi}_1, \cdots, \boldsymbol{\xi}_T\right), V\left(\boldsymbol{\xi}_1, \cdots, \boldsymbol{\xi}_T\right)\right)\right\}-E\left\{f\left(W\left(\boldsymbol{Y}_1, \cdots, \boldsymbol{Y}_T\right), V\left(\boldsymbol{Y}_1, \cdots, \boldsymbol{Y}_T\right)\right)\right\}\right| \\
	\leq & \sum_{d=1}^T \mid E\left\{f\left(W_d, V_d\right)\right\}-E\left\{f\left(W_{d+1}, V_{d+1}\right)\right\} \mid\\
	=& \sum_{d=1}^T \mid E \left[E\left\{f\left(W_d, V_d\right)| \mathcal{F}_{fr}\right\}-E\left\{f\left(W_{d+1}, V_{d+1}\right)| \mathcal{F}_{fr}\right\}\right]\mid\\
	\leq &\sum_{d=1}^T  E \mid E\left\{f\left(W_d, V_d\right)| \mathcal{F}_{fr}\right\}-E\left\{f\left(W_{d+1}, V_{d+1}| \mathcal{F}_{fr}\right)\right\}\mid
\end{align*}}
Let
\begin{align*}
	W_{d, 0} & =\frac{2 N\omega_T^{-1} \sum_{i<j}^{d-1} h_ih_j\boldsymbol{\xi}_i^{\top} \boldsymbol{\xi}_j+2 N\omega_T^{-1}\sum_{d+1 \leq i<j \leq T}h_ih_j \boldsymbol{Y}_i^{\top} \boldsymbol{Y}_j}{\sigma_{SS}}\\
	&\quad+\frac{2 N\omega_T^{-1} \sum_{i=1}^{d-1} \sum_{j=d+1}^T h_ih_j\boldsymbol{\xi}_i^{\top} \boldsymbol{Y}_j}{\sigma_{SS}} \in \mathcal{F}_d \\
	V_{d, 0}  &=\beta^{-1} \log \bigg(\sum_{j=1}^N\exp \bigg(\beta \sqrt{\frac{N}{T}} \sum_{i=1}^{d-1} (1-\sum_{s=1}^Tr_s^{-1}r_iV_{si}) \xi_{i, j}\\
	&\quad+\beta \sqrt{\frac{N}{T}} \sum_{i=d+1}^T (1-\sum_{s=1}^Tr_s^{-1}r_iV_{si}) Y_{i, j}\bigg)\bigg) \in \mathcal{F}_d
\end{align*}
By Taylor expansion, we have,
\begin{align*}
	&f\left(W_d, V_d\right)-f\left(W_{d, 0}, V_{d, 0}\right)\\
	= & f_1\left(W_{d, 0}, V_{d, 0}\right)\left(W_d-W_{d, 0}\right)+f_2\left(W_{d, 0}, V_{d, 0}\right)\left(V_d-V_{d, 0}\right) \\
	& +\frac{1}{2} f_{11}\left(W_{d, 0}, V_{d, 0}\right)\left(W_d-W_{d, 0}\right)^2+\frac{1}{2} f_{22}\left(W_{d, 0}, V_{d, 0}\right)\left(V_d-V_{d, 0}\right)^2 \\
	& +\frac{1}{2} f_{12}\left(W_{d, 0}, V_{d, 0}\right)\left(W_d-W_{d, 0}\right)\left(V_d-V_{d, 0}\right) \\
	& +O\left(\left|\left(V_d-V_{d, 0}\right)\right|^3\right)+O\left(\left|\left(W_d-W_{d, 0}\right)\right|^3\right),
\end{align*}
and
\begin{align*}
	&f\left(W_{d+1}, V_{d+1}\right)-f\left(W_{d, 0}, V_{d, 0}\right)\\
	= & f_1\left(W_{d, 0}, V_{d, 0}\right)\left(W_{d+1}-W_{d, 0}\right)+f_2\left(W_{d, 0}, V_{d, 0}\right)\left(V_{d+1}-V_{d, 0}\right) \\
	& +\frac{1}{2} f_{11}\left(W_{d, 0}, V_{d, 0}\right)\left(W_{d+1}-W_{d, 0}\right)^2+\frac{1}{2} f_{22}\left(W_{d, 0}, V_{d, 0}\right)\left(V_{d+1}-V_{d, 0}\right)^2 \\
	& +\frac{1}{2} f_{12}\left(W_{d, 0}, V_{d, 0}\right)\left(W_{d+1}-W_{d, 0}\right)\left(V_{d+1}-V_{d, 0}\right) \\
	& +O\left(\left|\left(V_{d+1}-V_{d, 0}\right)\right|^3\right)+O\left(\left|\left(W_{d+1}-W_{d, 0}\right)\right|^3\right),
\end{align*}
where for $f:=f(x, y)$, $f_1(x, y)=\frac{\partial f}{\partial x}, f_2(x, y)=\frac{\partial f}{\partial y}, f_{11}(x, y)=\frac{\partial f^2}{\partial^2 x}, f_{22}(x, y)=\frac{\partial f^2}{\partial^2 y}$ and $f_{12}(x, y)=\frac{\partial f^2}{\partial x \partial y}$.
We first consider $W_d, W_{d+1}, W_{d, 0}$ and notice that,
\begin{align*}
	W_d-W_{d, 0} & =\frac{N\omega_T^{-1} \sum_{i=1}^{d-1} h_ih_d\boldsymbol{\xi}_i^{\top} \boldsymbol{Y}_d+ N\omega_T^{-1}\sum_{i=d+1}^T h_ih_d\boldsymbol{Y}_i^{\top} \boldsymbol{Y}_d}{\sigma_{SS}}, \\
	W_{d+1}-W_{d, 0} & =\frac{N\omega_T^{-1} \sum_{i=1}^{d-1} h_ih_d\boldsymbol{\xi}_i^{\top} \boldsymbol{\xi}_d+N\omega_T^{-1}  \sum_{i=d+1}^T h_ih_d\boldsymbol{Y}_i^{\top} \boldsymbol{\xi}_d}{\sigma_{SS}} .
\end{align*}
Due to $E\left(\boldsymbol{\xi}_t\right)=E\left(\boldsymbol{Y}_t\right)=0$ and $E\left(\boldsymbol{\xi}_t \boldsymbol{\xi}_t^{\top}\right)=E\left(\boldsymbol{Y}_t \boldsymbol{Y}_t^{\top}\right)$, it can be verified that, $$E\left(W_d-W_{d, 0} \mid \mathcal{F}_d,\mathcal{F}_{fr}\right)=E\left(W_{d+1}-W_{d, 0} \mid \mathcal{F}_d,\mathcal{F}_{fr}\right)$$ and $E\left(\left(W_d-W_{d, 0}\right)^2 \mid \mathcal{F}_d,\mathcal{F}_{fr}\right)=E\left(\left(W_{d+1}-W_{d, 0}\right)^2 \mid \mathcal{F}_d,\mathcal{F}_{fr}\right)$.
Hence,
\begin{align*}
	E\left\{f_1\left(W_{d, 0}, V_{d, 0}\right)\left(W_d-W_{d, 0}\right)\mid\mathcal{F}_{fr}\right\} & =E\left\{f_1\left(W_{d, 0}, V_{d, 0}\right)\left(W_{d+1}-W_{d, 0}\right)\mid\mathcal{F}_{fr}\right\} \text { and } \\
	E\left\{f_{11}\left(W_{d, 0}, V_{d, 0}\right)\left(W_d-W_{d, 0}\right)^2\mid\mathcal{F}_{fr}\right\} & =E\left\{f_{11}\left(W_{d, 0}, V_{d, 0}\right)\left(W_{d+1}-W_{d, 0}\right)^2\mid\mathcal{F}_{fr}\right\} .
\end{align*}
Next we consider $V_d-V_{d, 0}$. Let $z_{d, 0, j}=\sqrt{\frac{N}{T}} \sum_{i=1}^{d-1} (1-\sum_{s=1}^Tr_s^{-1}r_iV_{si})\xi_{i, j}+\sqrt{\frac{N}{T}} \sum_{i=d+1}^T (1-\sum_{s=1}^Tr_s^{-1}r_iV_{si})Y_{i, j}$, $z_{d, j}=z_{d, 0, j}+T^{-1 / 2} \sqrt{N} (1-\sum_{s=1}^Tr_s^{-1}r_dV_{sd})Y_{d, j}$, $z_{d+1, j}=z_{d, 0, j}+T^{-1 / 2} \sqrt{N} (1-\sum_{s=1}^Tr_s^{-1}r_dV_{sd})\xi_{d, j}$. By Taylor expansion, we have that:
\begin{align*}
	&V_d-V_{d, 0}\\
	= & \sum_{l=1}^N \partial_l F_\beta\left(\boldsymbol{z}_{d, 0}\right)\left(z_{d, l}-z_{d, 0, l}\right)+\frac{1}{2} \sum_{l=1}^N \sum_{k=1}^N \partial_k \partial_l F_\beta\left(\boldsymbol{z}_{d, 0}\right)\left(z_{d, l}-z_{d, 0, l}\right)\left(z_{d, k}-z_{d, 0, k}\right) \\
	& +\frac{1}{6} \sum_{l=1}^N \sum_{k=1}^N \sum_{v=1}^N \partial_v \partial_k \partial_l F_\beta\left(\boldsymbol{z}_{d, 0}+\delta\left(\boldsymbol{z}_d-\boldsymbol{z}_{d, 0}\right)\right)\left(z_{d, l}-z_{d, 0, l}\right)\left(z_{d, k}-\boldsymbol{z}_{d, 0, k}\right)\left(\boldsymbol{z}_{d, v}-\boldsymbol{z}_{d, 0, v}\right),
\end{align*}
for some $\delta \in(0,1)$. Again, due to $E\left(\boldsymbol{\xi}_t\right)=E\left(\boldsymbol{Y}_t\right)=0$ and $E\left(\boldsymbol{\xi}_t \boldsymbol{\xi}_t^{\top}\right)=E\left(\boldsymbol{Y}_t \boldsymbol{Y}_t^{\top}\right)$, we can verify that
$E\left\{\left(z_{d, l}-z_{d, 0, l}\right) \mid \mathcal{F}_d,\mathcal{F}_{fr}\right\}=E\left\{\left(z_{d+1, l}-z_{d, 0, l}\right) \mid \mathcal{F}_d,\mathcal{F}_{fr}\right\}$ and $$E\left\{\left(z_{d, l}-z_{d, 0, l}\right)^2 \mid \mathcal{F}_d,\mathcal{F}_{fr}\right\}=E\left\{\left(z_{d+1, l}-z_{d, 0, l}\right)^2 \mid \mathcal{F}_d,\mathcal{F}_{fr}\right\}.$$
By Lemma A.2 in \cite{chernozhukov2013gaussian}, we have,
$$
\left|\sum_{l=1}^N \sum_{k=1}^N \sum_{v=1}^N \partial_v \partial_k \partial_l F_\beta\left(\boldsymbol{z}_{d, 0}+\delta\left(\boldsymbol{z}_d-\boldsymbol{z}_{d, 0}\right)\right)\right| \leq C \beta^2,
$$
for some positive constant $C$.
By Lemma A.4 in \cite{cheng2023}, we have that: $\left\|\zeta_1^{-1} U_{i, j}\right\|_{\psi_\alpha} \lesssim \bar{B}$, for all $i=1, \ldots, T$ and $j=1, \ldots, N$, which means $$P\left(\left|\sqrt{N} \xi_{i, j}\right| \geq t\right) \leq 2 \exp \left(-\left(c t  / \{\zeta_1\sqrt{N}\}\right)^\alpha\right) \lesssim 2 \exp \left(-(c t)^\alpha\right),$$ ${P}\left(\max _{1 \leq j \leq N}\left|\sqrt{N} \xi_{i j}\right|>C \log N\right) \rightarrow$ 0 and since $\sqrt{N} Y_{t j} \sim N(0,1)$, $${P}\left(\max _{1 \leq j \leq N}\left|\sqrt{N} Y_{i j}\right|>C \log N\right) \rightarrow 0.$$ Hence,
{ \begin{align*}
	& \left|\frac{1}{6} \sum_{l=1}^N \sum_{k=1}^N \sum_{v=1}^N \partial_v \partial_k \partial_l F_\beta\left(\boldsymbol{z}_{d, 0}+\delta\left(\boldsymbol{z}_d-\boldsymbol{z}_{d, 0}\right)\right)\left(\boldsymbol{z}_{d, l}-\boldsymbol{z}_{d, 0, l}\right)\left(\boldsymbol{z}_{d, k}-\boldsymbol{z}_{d, 0, k}\right)\left(\boldsymbol{z}_{d, v}-\boldsymbol{z}_{d, 0, v}\right)\right| \\
	& \leq C \beta^2 |1-\sum_{s=1}^Tr_s^{-1}r_dV_{sd}|^3T^{-3 / 2} \log ^3(N), \\
	& \Big|\frac{1}{6} \sum_{l=1}^N \sum_{k=1}^N \sum_{v=1}^N \partial_v \partial_k \partial_l F_\beta\left(\boldsymbol{z}_{d+1,0}+\delta\left(\boldsymbol{z}_{d+1}-\boldsymbol{z}_{d, 0}\right)\right)\left(\boldsymbol{z}_{d+1, l}-\boldsymbol{z}_{d, 0, l}\right)\left(\boldsymbol{z}_{d+1, k}-\boldsymbol{z}_{d, 0, k}\right)\\
	&\quad\quad\times\left(\boldsymbol{z}_{d+1, v}-\boldsymbol{z}_{d, 0, v}\right)\Big| \\
	& \leq C \beta^2 |1-\sum_{s=1}^Tr_s^{-1}r_dV_{sd}|^3T^{-3 / 2} \log ^3(N),
\end{align*}}
holds with probability approaching one. Consequently, we have that: with probability one,
{ \begin{align*}
&\left|E\left\{f_2\left(W_{d, 0}, V_{d, 0}\right)\left(V_d-V_{d, 0}\right)\mid\mathcal{F}_{fr}\right\}-E\left\{f_2\left(W_{d, 0}, V_{d, 0}\right)\left(V_{d+1}-V_{d, 0}\right)\mid\mathcal{F}_{fr}\right\}\right| \\
\leq &C \beta^2 |1-\sum_{s=1}^Tr_s^{-1}r_dV_{sd}|^3T^{-3 / 2} \log ^3(N) .
\end{align*}
Similarly, it can be verified that,
\begin{align*}
&\left|E\left\{f_{22}\left(W_{d, 0}, V_{d, 0}\right)\left(V_d-V_{d, 0}\right)^2\mid\mathcal{F}_{fr}\right\}-E\left\{f_{22}\left(W_{d, 0}, V_{d, 0}\right)\left(V_{d+1}-V_{d, 0}\right)^2\mid\mathcal{F}_{fr}\right\}\right| \\
\leq &C \beta^2 |1-\sum_{s=1}^Tr_s^{-1}r_dV_{sd}|^3T^{-3 / 2} \log ^3(N),
\end{align*}
and
\begin{align*}
	& |E\left\{f_{12}\left(W_{d, 0}, V_{d, 0}\right)\left(W_d-W_{d, 0}\right)\left(V_d-V_{d, 0}\right)\mid\mathcal{F}_{fr}\right\}\\
	&-E\left\{f_{12}\left(W_{d, 0}, V_{d, 0}\right)\left(W_{d+1}-W_{d, 0}\right)\left(V_{d+1}-V_{d, 0}\right)\mid\mathcal{F}_{fr}\right\}| \\
	& \quad \leq C \beta^2 |1-\sum_{s=1}^Tr_s^{-1}r_dV_{sd}|^3T^{-3 / 2} \log ^3(N) .
\end{align*}
Hence, we have $E\left(\left|V_d-V_{d, 0}\right|^3\mid\mathcal{F}_{fr}\right)=O\Big\{|1-\sum_{s=1}^Tr_s^{-1}r_dV_{sd}|^3T^{-3 / 2} \log ^3(N)\Big\}$.}

 For $E\left(\left|W_d-W_{d, 0}\right|^3\mid\mathcal{F}_{fr}\right)$, we first calculate $E\left(\left(W_d-W_{d, 0}\right)^4\mid\mathcal{F}_{fr}\right)$, then it's easy to get the order for 3-order term.
\begin{align*}
	E\left(\left(W_d-W_{d, 0}\right)^4\mid\mathcal{F}_{fr}\right) & =E\left(\frac{N\omega_T^{-1} \sum_{i=1}^{d-1} h_ih_d\boldsymbol{\xi}_i^{\top} \boldsymbol{Y}_d+N\omega_T^{-1} \sum_{i=d+1}^T h_ih_d \boldsymbol{Y}_i^{\top} \boldsymbol{Y}_d}{\sigma_{SS}}\right)^4 \\
	& =\frac{N^{4}\omega_T^{-4}}{4\left(\operatorname{tr}\left(\mathbf{R}^2\right)\right)^2} E\left(\sum_{i=1}^{d-1} h_ih_d\boldsymbol{\xi}_i^{\top} \boldsymbol{Y}_d+\sum_{i=d+1}^T h_ih_d\boldsymbol{Y}_i^{\top} \boldsymbol{Y}_d\right)^4 .
\end{align*}
We consider the binomial expansion term and calculate them separately in above equation:
\begin{align}\label{7.19}
	(i) & =E_1\left(\sum_{i=d+1}^T h_ih_d\boldsymbol{Y}_i^{\top} \boldsymbol{Y}_d\right)^4,(i i)=E_1 \sum_{i=1}^{d-1} h_ih_d\boldsymbol{\xi}_i^{\top} \boldsymbol{Y}_d \left(\sum_{i=d+1}^Th_ih_d \boldsymbol{Y}_i^{\top} \boldsymbol{Y}_d\right)^3,\n\\
	(i i i)&=E_1\left(\sum_{i=1}^{d-1} h_ih_d\boldsymbol{\xi}_i^{\top} \boldsymbol{Y}_d\right)^2 \left(\sum_{i=d+1}^T \boldsymbol{Y}_i^{\top}h_ih_d \boldsymbol{Y}_d\right)^2, \n\\
	(i v) &=E_1\left(\sum_{i=1}^{d-1}  h_ih_d\boldsymbol{\xi}_i^{\top} \boldsymbol{Y}_d\right)^3  \sum_{i=d+1}^T  h_ih_d\boldsymbol{Y}_i^{\top} \boldsymbol{Y}_d,
	(v)=E_1\left(\sum_{i=1}^{d-1} h_ih_d\boldsymbol{\xi}_i^{\top} \boldsymbol{Y}_d\right)^4,
\end{align}
where $E_1$ means conditional expectation given  $\mathcal{F}_{fr}$.
Since $E \boldsymbol{Y}_i=E \boldsymbol{\xi}_i=0$, we easily find that \eqref{7.19}-(ii)(iv) equal to 0. Next we can get the following equations for \eqref{7.19}-(iii) after some straightforward calculations.
\begin{align*}
	&\quad E_1\left(\sum_{i=1}^{d-1} h_ih_d\boldsymbol{\xi}_i^{\top} \boldsymbol{Y}_d\right)^2 \left(\sum_{i=d+1}^T h_ih_d\boldsymbol{Y}_i^{\top} \boldsymbol{Y}_d\right)^2 \\
	& =E_1\left[E_1\left[\left(\sum_{i=1}^{d-1}h_ih_d \boldsymbol{\xi}_i^{\top} \boldsymbol{Y}_d\right)^2\left(\sum_{i=d+1}^Th_ih_d \boldsymbol{Y}_i^{\top} \boldsymbol{Y}_d\right)^2 \mid \boldsymbol{Y}_d\right]\right] \\
	& =E_1\left[\sum_{i=1}^{d-1}\sum_{j=d+1}^{T}h_i^2h_j^2h_d^4\left(\boldsymbol{Y}_d^{\top} \boldsymbol{\Sigma}_u \boldsymbol{Y}_d\right)^2\right] \\
	& =E_1\left[\sum_{i=1}^{d-1}\sum_{j=d+1}^{T}h_i^2h_j^2h_d^4\left(\left(\boldsymbol{\Sigma}_u^{-1 / 2} \boldsymbol{Y}_d\right)^{\top} \boldsymbol{\Sigma}_u^2\left(\boldsymbol{\Sigma}_u^{-1 / 2} \boldsymbol{Y}_d\right)\right)^2\right] \\
	& =\sum_{i=1}^{d-1}\sum_{j=d+1}^{T}h_i^2h_j^2h_d^4 \cdot 2 \operatorname{tr}\left(\boldsymbol{\Sigma}_u^4\right) \\
	& \leq\sum_{i=1}^{d-1}\sum_{j=d+1}^{T}h_i^2h_j^2h_d^4 \cdot O\left(\operatorname{tr}^2\left(\boldsymbol{\Sigma}_u^2\right)\right) .
\end{align*}

By some properties for standard normal random variable, the last inequality holds with some simple calculations shown below: (i) $\operatorname{tr}\left(\boldsymbol{\Sigma}_u^4\right) =\left\|\boldsymbol{\Sigma}_u^2\right\|_F^2=\left\|\boldsymbol{\Sigma}_u \cdot \boldsymbol{\Sigma}_u\right\|_F^2 \leq\operatorname{tr}^2\left(\boldsymbol{\Sigma}_u^2\right) .$
(ii) If $\boldsymbol{X}, \boldsymbol{Y} \stackrel{\text { i.i.d. }}{\sim} N\left(0, \mathbf{I}_p\right)$, then
\begin{align*}
	E_1\left(\boldsymbol{X}^{\top} \mathbf{A} \boldsymbol{X}\right)^2 & =2 \operatorname{tr}\left(\mathbf{A}^2\right)+\operatorname{tr}^2(\mathbf{A}) \leq2 \operatorname{tr}^2\left(\mathbf{A}\right), \\
	E_1\left(\boldsymbol{X}^{\top} \mathbf{A} \boldsymbol{Y}\right)^4 & =E_1\left[E_1\left[\left(\boldsymbol{Y}^{\top} \mathbf{A} \boldsymbol{X} \boldsymbol{X}^{\top} \mathbf{A} \boldsymbol{Y}\right)^2 \mid \boldsymbol{X}\right]\right] \leq 2 E_1\left[\operatorname{tr}^2\left(\mathbf{A} \boldsymbol{X} \boldsymbol{X}^{\top} \mathbf{A}\right)\right] \\
	& =2 E_1\left[\left(\boldsymbol{X}^{\top} \mathbf{A}^2 \boldsymbol{X}\right)^2\right] \leq 4 \operatorname{tr}^2\left(\mathbf{A}^2\right) .
\end{align*}
For \eqref{7.19}-(i), according to $\sum_{i=d+1}^T h_i\boldsymbol{Y}_i \sim N\left(0,\sum_{i=d+1}^T h_i^2\boldsymbol{\Sigma}_u\right)$ and above inequalities, we have,
\begin{align*}
	&E_1\left(\sum_{i=d+1}^T h_ih_d\boldsymbol{Y}_i^{\top} \boldsymbol{Y}_d\right)^4 \\ =&E_1\left(\left(\frac{1}{\sqrt{\sum_{i=d+1}^T h_i^2}} \boldsymbol{\Sigma}_u^{-1 / 2} \sum_{i=d+1}^T h_i \boldsymbol{Y}_i\right)^{\top}\left(\sqrt{\sum_{i=d+1}^T h_i^2h_d^2} \boldsymbol{\Sigma}_u\right)\left(\boldsymbol{\Sigma}_u^{-1 / 2} \boldsymbol{Y}_d\right)\right)^4 \\
	 \leq&\left(\sum_{i=d+1}^T h_i^2h_d^2\right)^2 O\left(\operatorname{tr}^2\left(\boldsymbol{\Sigma}_u^2\right)\right)
\end{align*}
Similarly, for \eqref{7.19}-(v),
$$
E_1\left(\sum_{i=1}^{d-1} h_ih_d\boldsymbol{\xi}_i^{\top} \boldsymbol{Y}_d\right)^4 \leq\left(\sum_{i=1}^{d-1} h_i^2\right)^2h_d^4 O\{\operatorname{tr}^2\left(\boldsymbol{\Sigma}_u^2\right) \}
$$

Thus, combined with above results,
\begin{align*}
	& E_1\left(\left(W_d-W_{d, 0}\right)^4\right) \\
	= & \frac{N^4\omega_T^{-4}}{4\left\{\operatorname{tr}\left(\R^2\right)\right\}^2} E_1\left(\sum_{i=1}^{d-1} h_ih_d\boldsymbol{\xi}_i^{\top} \boldsymbol{Y}_d+\sum_{i=d+1}^T h_ih_d\boldsymbol{Y}_i^{\top} \boldsymbol{Y}_d\right)^4 \\
	\leq & \frac{N^4\omega_T^{-4}}{4\left\{\operatorname{tr}\left(\R^2\right)\right\}^2}\left\{6\left(\sum_{i=1}^{d-1}h_i^2\right)\left(\sum_{i=d+1}^{T-d}h_i^2\right)+\left(\sum_{i=d+1}^{T-d}h_i^2\right)^2+\left(\sum_{i=1}^{d-1}h_i^2\right)^2\right\} O\left(h_d^4\operatorname{tr}^2\left(\boldsymbol{\Sigma}_u^2\right)\right) \\
	\leq & \frac{N^4\omega_T^{-4}}{4\left\{\operatorname{tr}\left(\R^2\right)\right\}^2} O(\omega_T^2) O\left(\operatorname{tr}\left(\boldsymbol{\Sigma}_u^2\right)^2\right)=O\left(\frac{1}{T^2}\right) .
\end{align*}
By Jensen's inequality, we get
$$
\sum_{d=1}^T E_1\left|W_d-W_{d, 0}\right|^3 \leq \sum_{d=1}^T\left(E_1\left(W_d-W_{d, 0}\right)^4\right)^{3 / 4} \leq C^{\prime} T^{-1 / 2},
$$
for some positive constant $C^{\prime}$, Combining all facts together, we conclude that
{ \begin{align*}
&\sum_{d=1}^T  E \mid E\left\{f\left(W_d, V_d\right)| \mathcal{F}_{fr}\right\}-E\left\{f\left(W_{d+1}, V_{d+1}| \mathcal{F}_{fr}\right)\right\}\mid \\
\leq& C\beta^2E\sum_{d=1}^T |1-\sum_{s=1}^Tr_s^{-1}r_dV_{sd}|^3 T^{-3 / 2} \log ^3(N)+C^{\prime} T^{-1 / 2}\\
\leq &C\beta^2O(1)T^{-1 / 2} \log ^3(N)+C^{\prime} T^{-1 / 2} \rightarrow 0,
\end{align*}}
as $(N, T) \rightarrow \infty$. The conclusion follows.
\hfill$\Box$
\subsection{Proof of Theorem 4}
From the proof of Theorem 2 in \cite{liu2023high}, we can find that
$$
T_{S S}=\frac{N(\h\trans\h)^{-1}\sum \sum_{t_1\neq t_2} h_{t_1}h_{t_2}\boldsymbol{U}_{t_1}^{\top} \boldsymbol{U}_{t_2}+N(\h\trans\h-1)\xi_1^2 \boldsymbol{\alpha}^{\top} \mathbf{D}^{-1} \boldsymbol{\alpha}}{\sqrt{2\tr^2(\R^2)}}+o_p\left(1\right),
$$
and according to \eqref{eq4}, we get the Bahadur representation in $L^{\infty}$ norm,
$$
T^{1 / 2} \mathbf{D}^{-1 / 2}\hat{\boldsymbol{\theta}}=T^{-1 / 2} \zeta_1^{-1} \sum_{i=1}^T\Big\{(1-\sum_{s=1}^Tr_s^{-1}r_iV_{si})\boldsymbol{U}_i+\omega{\bm\alpha}\Big\}+C_T .
$$
Similar with the proof in Theorem 3, it's suffice to show the result holds for normal version, i.e. it suffice to show that:
$$
\left\|N^{1/2} \sum_{t=1}^T\omega_{T}^{-1/2} h_{t}\boldsymbol{Y}_t\right\|^2 \text { and }\left\|\sqrt{\frac{N}{T}} \sum_{i=1}^T\Big\{(1-\sum_{s=1}^Tr_s^{-1}r_iV_{si})\boldsymbol{Y}_i+\zeta_1 \mathbf{D}^{-1 / 2} \omega\boldsymbol{\alpha}\Big\}\right\|_{\infty}^2,
$$
are asymptotic independent, where $\left\{\boldsymbol{Y}_1, \boldsymbol{Y}_2, \cdots, \boldsymbol{Y}_T\right\}$ are sample from $N\left(0, E \boldsymbol{U}_1^{\top} \boldsymbol{U}_1\right)$ and $\boldsymbol{Y}_t=(y_{1t},\dots,y_{Nt})$.
Denote $N^{1/2} \sum_{t=1}^T\omega_{T}^{-1/2} h_{t}\boldsymbol{Y}_t:=\boldsymbol{\varphi}=\left(\varphi_1, \cdots, \varphi_N\right)^{\top}, \boldsymbol{\varphi}_{\mathcal{A}}=\left(\varphi_{j_1}, \cdots, \varphi_{j_d}\right)^{\top}$, and $\boldsymbol{\varphi}_{\mathcal{A}^c}=\left(\varphi_{j_{d+1}}, \cdots, \varphi_{j_N}\right)^{\top}$, where $\mathcal{A}=\left\{j_1, j_2, \cdots, j_d\right\}$ is the set of $\{i:\alpha_i\neq 0\}$ and $d=|\mathcal{A}|$. Then, $S=\|\boldsymbol{\varphi}\|^2=\left\|\boldsymbol{\varphi}_{\mathcal{A}}\right\|^2+\left\|\boldsymbol{\varphi}_{\mathcal{A}^c}\right\|^2$,
 \begin{align*}
M=&\left\|\sqrt{\frac{N}{T}} \sum_{t=1}^T\boldsymbol{Y}_t+\sqrt{N T^{-1}} \zeta_1 \mathbf{D}^{-1 / 2} \omega\bm \alpha\right\|_{\infty}\\
=&\max\bigg\{\max _{i \in \mathcal{A}}\left(\sqrt{\frac{N}{T}} \sum_{t=1}^T(1-\sum_{s=1}^Tr_s^{-1}r_tV_{st}) {y}_{it}+\sqrt{N T^{-1}} \zeta_1 \mathbf{D}^{-1 / 2} \omega\bm \alpha\right),\\
&\quad\quad\quad\max _{i \in \mathcal{A}^c} \sqrt{\frac{N}{T}} \sum_{t=1}^T (1-\sum_{s=1}^Tr_s^{-1}r_tV_{st}){y}_{it}\bigg\}.
\end{align*}
 From the proof of Theorem 6, we know that $\left\|\boldsymbol{\varphi}_{\mathcal{A}^c}\right\|^2$ and $$\max _{i \in \mathcal{A}^c} \sqrt{NT^{-1}} \sum_{t=1}^T(1-\sum_{s=1}^Tr_s^{-1}r_tV_{st}) {y}_{it}$$ are asymptotically independent. Hence, it suffice to show that $\left\|\boldsymbol{\varphi}_{\mathcal{A}^c}\right\|^2$ is asymptotically independent with $\max _{i \in \mathcal{A}}\sqrt{NT^{-1}} \sum_{t=1}^T (1-\sum_{s=1}^Tr_s^{-1}r_tV_{st}){y}_{it}$. Hence, we just need to prove that $\left\|\boldsymbol{\varphi}_{\mathcal{A}^c}\right\|^2$ is asymptotically independent with  $\max _{i \in \mathcal{A}}\sqrt{N/T} \sum_{t=1}^T (1-\sum_{s=1}^Tr_s^{-1}r_tV_{st}){y}_{it}$.
 
 { 
  Obviously, we have $E(\boldsymbol{\varphi}_{\mathcal{A}^c}\boldsymbol{\varphi}_{\mathcal{A}^c}^{\top})=N\mathbf{\Sigma}_{u,\mathcal{A}^c,\mathcal{A}^c}$. Define $\Y_{t,\mathcal{A}}=(y_{j_1t},\dots,y_{j_dt})$ for any $1\leq t\leq T$. Then, we have $E(\sqrt{NT^{-1}}\sum_{t=1}^{T} (1-\sum_{s=1}^Tr_s^{-1}r_tV_{st})\Y_{t,\mathcal{A}}\sqrt{NT^{-1}}\sum_{t=1}^{T} (1-\sum_{s=1}^Tr_s^{-1}r_tV_{st})\Y_{t,\mathcal{A}}^{\top})=NT^{-1}\sum_{t=1}^{T} E(1-\sum_{s=1}^Tr_s^{-1}r_tV_{st})^2\mathbf{\Sigma}_{u,\mathcal{A},\mathcal{A}}=O(N\mathbf{\Sigma}_{u,\mathcal{A},\mathcal{A}})$ and $$E(\boldsymbol{\varphi}_{\mathcal{A}^c}\sqrt{NT^{-1}}\sum_{t=1}^{T}\Y_{t,\mathcal{A}}^{\top})=N\sqrt{T^{-1}}\sum_{t=1}^{T}E\Big\{\omega_T^{-1/2}h_t(1-\sum_{s=1}^Tr_s^{-1}r_tV_{st})\Big\}\mathbf{\Sigma}_{u,\mathcal{A}^c,\mathcal{A}}=O(N\mathbf{\Sigma}_{u,\mathcal{A}^c,\mathcal{A}}).$$ 
 Let $\mathbf{\Sigma}_{22}=NT^{-1}\sum_{t=1}^{T} E(1-\sum_{s=1}^Tr_s^{-1}r_tV_{st})^2\mathbf{\Sigma}_{u,\mathcal{A},\mathcal{A}}$, $\mathbf{\Sigma}_{21}=N\sqrt{T^{-1}}\sum_{t=1}^{T}E\Big\{\omega_T^{-1/2}h_t(1-\sum_{s=1}^Tr_s^{-1}r_tV_{st})\Big\}\mathbf{\Sigma}_{u,\mathcal{A},\mathcal{A}^c}$
  and $\mathbf{\Sigma}_{12}=N\sqrt{T^{-1}}\sum_{t=1}^{T}E\Big\{\omega_T^{-1/2}h_t(1-\sum_{s=1}^Tr_s^{-1}r_tV_{st})\Big\}\mathbf{\Sigma}_{u,\mathcal{A}^c,\mathcal{A}}.$ 
  Define \begin{align*}
 	\mathbf{\Sigma}_{U}=\left(\begin{array}{cc}
 	N\mathbf{\Sigma}_{u,\mathcal{A}^c,\mathcal{A}^c} & \mathbf{\Sigma}_{12} \\
 \mathbf{\Sigma}_{21} & \mathbf{\Sigma}_{22}
 \end{array}\right).
 \end{align*}
 By Theorem 1.2.11 in \cite{muirhead2009aspects}, $\boldsymbol{\varphi}_{\mathcal{A}^c}$ can be decomposed as $\boldsymbol{\varphi}_{\mathcal{A}^c}=\boldsymbol{E}+\boldsymbol{F}$, where $\boldsymbol{E}=\boldsymbol{\varphi}_{\mathcal{A}^c}-\boldsymbol{\Sigma}_{U, \mathcal{A}^c, \mathcal{A}} \boldsymbol{\Sigma}_{U, \mathcal{A}, \mathcal{A}}^{-1} \boldsymbol{\varphi}_{\mathcal{A}}, \boldsymbol{F}=$ $\boldsymbol{\Sigma}_{U, \mathcal{A}^c, \mathcal{A}} \boldsymbol{\Sigma}_{U, \mathcal{A}, \mathcal{A}}^{-1} \boldsymbol{\varphi}_{\mathcal{A}}$, which fulfill the properties $\boldsymbol{E} \sim N\left(0, \boldsymbol{\Sigma}_{U, \mathcal{A}^c, \mathcal{A}^c}-\right.$ $\left.\boldsymbol{\Sigma}_{U, \mathcal{A}^c, \mathcal{A}} \boldsymbol{\Sigma}_{U, \mathcal{A}, \mathcal{A}}^{-1} \boldsymbol{\Sigma}_{U, \mathcal{A}, \mathcal{A}^c}\right), \boldsymbol{F} \sim N\left(0, \boldsymbol{\Sigma}_{U, \mathcal{A}^c, \mathcal{A}} \boldsymbol{\Sigma}_{U, \mathcal{A}, \mathcal{A}}^{-1} \boldsymbol{\Sigma}_{U, \mathcal{A}, \mathcal{A}^c}\right)$ and $\boldsymbol{E}$ and $\boldsymbol{\varphi}_{\mathcal{A}}$ are independent.}
Then, we rewrite
$$
\left\|\boldsymbol{\varphi}_{\mathcal{A}^c}\right\|^2=\boldsymbol{E}^{\top} \boldsymbol{E}+\boldsymbol{F}^{\top} \boldsymbol{F}+2 \boldsymbol{E}^{\top} \boldsymbol{F}.
$$
According the proof of lemma S.7 in \cite{feng2022asymptotic}, we have that:
$$
P\left(\left|\boldsymbol{F}^{\top} \boldsymbol{F}+2 \boldsymbol{E}^{\top} \boldsymbol{F}\right| \geq \epsilon \left[2 \operatorname{tr}\left(\mathbf{R}^2\right)\right]^{1 / 2}\right) \leq \frac{3}{N^t} \rightarrow 0,
$$
by $d=o\left(\lambda_{\min }(\mathbf{R}) [\operatorname{tr}\left(\mathbf{R}^2\right)]^{1 / 2} /(\log N)^C\right)$, where $$t=t_N:=C \epsilon / 8 \cdot \left[2 \operatorname{tr}\left(\mathbf{R}^2\right)\right]^{1 / 2}/\left[\lambda_{\max }(\mathbf{R}) \log N\right] \rightarrow\infty,$$ and $\epsilon_N:=(\log N)^C /\left(\left[2 \operatorname{tr}\left(\mathbf{R}^2\right)\right]^{1 / 2} \lambda_{\min }(\mathbf{R})\right)\rightarrow 0$.
\hfill$\Box$

\end{document}